\documentclass[longauth]{aa}  

\usepackage{graphicx}
\usepackage{txfonts}
\usepackage{amsmath}
\usepackage{amssymb}
\usepackage{gensymb}
\usepackage{flexisym}
\usepackage{natbib}
\bibpunct{(}{)}{;}{a}{}{,} 

\usepackage[switch]{lineno}

\begin{document} 

   \title{Broadband characterisation of the very intense TeV flares of the blazar 1ES 1959+650 in 2016}
   \titlerunning{Broadband characterisation of the very intense TeV flares of the blazar 1ES 1959+650 in 2016}
%
\authorrunning{V.~A.~Acciari et al.}
\author{
MAGIC Collaboration:
V.~A.~Acciari\inst{1} \and
S.~Ansoldi\inst{2,21} \and
L.~A.~Antonelli\inst{3} \and
A.~Arbet Engels\inst{4} \and
D.~Baack\inst{5} \and
A.~Babi\'c\inst{6} \and
B.~Banerjee\inst{7} \and
U.~Barres de Almeida\inst{8} \and
J.~A.~Barrio\inst{9}
% ...
}
 
\institute { Inst. de Astrof\'isica de Canarias, E-38200 La Laguna, and Universidad de La Laguna, Dpto. Astrof\'isica, E-38206 La Laguna, Tenerife, Spain
\and Universit\`a di Udine, and INFN Trieste, I-33100 Udine, Italy
\and National Institute for Astrophysics (INAF), I-00136 Rome, Italy
\and ETH Zurich, CH-8093 Zurich, Switzerland
\and Technische Universit\"at Dortmund, D-44221 Dortmund, Germany
\and Croatian Consortium: University of Rijeka, Department of Physics, 51000 Rijeka; University of Split - FESB, 21000 Split; University of Zagreb - FER, 10000 Zagreb; University of Osijek, 31000 Osijek; Rudjer Boskovic Institute, 10000 Zagreb, Croatia
\and Saha Institute of Nuclear Physics, HBNI, 1/AF Bidhannagar, Salt Lake, Sector-1, Kolkata 700064, India
\and Centro Brasileiro de Pesquisas F\'isicas (CBPF), 22290-180 URCA, Rio de Janeiro (RJ), Brasil
\and Unidad de Part\'iculas y Cosmolog\'ia (UPARCOS), Universidad Complutense, E-28040 Madrid, Spain
\and University of \L\'od\'z, Department of Astrophysics, PL-90236 \L\'od\'z, Poland
\and Universit\`a di Siena and INFN Pisa, I-53100 Siena, Italy
\and Deutsches Elektronen-Synchrotron (DESY), D-15738 Zeuthen, Germany
\and Istituto Nazionale Fisica Nucleare (INFN), 00044 Frascati (Roma) Italy
\and Max-Planck-Institut f\"ur Physik, D-80805 M\"unchen, Germany
\and Institut de F\'isica d'Altes Energies (IFAE), The Barcelona Institute of Science and Technology (BIST), E-08193 Bellaterra (Barcelona), Spain
\and Universit\`a di Padova and INFN, I-35131 Padova, Italy
\and Universit\`a di Pisa, and INFN Pisa, I-56126 Pisa, Italy
\and Universit\"at W\"urzburg, D-97074 W\"urzburg, Germany
\and Finnish MAGIC Consortium: Tuorla Observatory (Department of Physics and Astronomy) and Finnish Centre of Astronomy with ESO (FINCA), University of Turku, FI-20014 Turku, Finland; Astronomy Division, University of Oulu, FI-90014 Oulu, Finland
\and Departament de F\'isica, and CERES-IEEC, Universitat Aut\`onoma de Barcelona, E-08193 Bellaterra, Spain
\and Japanese MAGIC Consortium: ICRR, The University of Tokyo, 277-8582 Chiba, Japan; Department of Physics, Kyoto University, 606-8502 Kyoto, Japan; Tokai University, 259-1292 Kanagawa, Japan; RIKEN, 351-0198 Saitama, Japan
% ...
}

   \date{Received xxx; accepted xxx}
   
   \abstract{1ES 1959+650 is a bright TeV high-frequency-peaked BL Lac object exhibiting interesting features like ``orphan'' TeV flares and a broad emission in the high-energy regime, that are difficult to interpret using conventional one-zone Synchrotron Self-Compton (SSC) scenarios. We report the results from the Major Atmospheric Gamma Imaging Cherenkov (MAGIC) observations in 2016 along with the multi-wavelength data from the \textit{Fermi} Large Area Telescope (LAT) and \textit{Swift} instruments. MAGIC observed 1ES 1959+650 with different emission levels in the very-high-energy (VHE, E $>$ 100 GeV) $\gamma$-ray band during 2016. In the long-term data, the X-ray spectrum becomes harder with increasing flux and a hint of a similar trend is also visible in the VHE band. An exceptionally high VHE flux reaching $\sim$3 times the Crab Nebula flux was measured by MAGIC on the 13th, 14th of June and 1st July 2016 (the highest flux observed since 2002). During these flares, the high-energy peak of the spectral energy distribution (SED) lies in the VHE domain and extends up to several TeV. 
   The spectrum in the $\gamma$-ray (both \textit{Fermi}-LAT and VHE bands) and the X-ray bands are quite hard.
   On 13th June and 1st July 2016, the source showed rapid variations of the VHE flux within timescales of less than an hour.
   A simple one-zone SSC model can describe the data during the flares requiring moderate to high values of the Doppler factors ($\delta \geq 30-60$).
   Alternatively, the high-energy peak of the SED can be explained by a purely hadronic model attributed to proton-synchrotron radiation with jet power $L_{jet}\sim10^{46}$ erg/s and under high values of the magnetic field strength ($\sim 100$ G) and maximum proton energy ($\sim$few EeV). Mixed lepto-hadronic models require super-Eddington values of the jet power.  
   We conclude that it is difficult to get detectable neutrino emission from the source during the extreme VHE flaring period of 2016.}


   \maketitle
%

\section{Introduction}
\label{sec:Introduction}

Blazars \citep{UrryPadovani} are a sub-class of active galactic nuclei (AGNs) that exhibit relativistic jets closely aligned to the line of sight of an observer on Earth. Due to strong relativistic beaming effects in this geometry, the intensity from these sources appears greatly boosted in the observer frame and is dominated by the non-thermal continuum produced within the jet. Blazars are characterised by rapid variability across the entire non-thermal waveband that spans over a wide energy range from radio to very-high-energy (VHE, $E>100$ GeV) $\gamma$ rays. 
BL~Lacs are a special class of blazar showing extremely weak or no emission lines in their optical/ultraviolet (UV) spectra.
\paragraph{}
Multi-wavelength (MWL) observations show that the non-thermal emission spectral energy distribution (SED) of blazars usually exhibits a double-peaked structure (e.g. \citealt{Pian1998}). The first SED peak is commonly attributed to synchrotron radiation of relativistic electrons located inside an emitting region within the jet and moving relativistically towards the observer with bulk Lorentz factor $\Gamma_{bulk}$. The origin of the high-energy peak in the SED is debatable. Within the framework of leptonic models, the origin of this component is often ascribed to Inverse Compton (IC) up-scattering of low-energy photons by high-energy electrons. 
For the BL Lacs the synchrotron photons present within the jet are commonly believed to serve as seeds for IC up-scattering (the so-called Synchrotron Self-Compton, SSC, scenario; \citealt{SSC,maraschi1992,dermer1993,GhiselliniIC}). Alternatively, in hadronic scenarios the second SED peak is attributed to relativistic protons accelerated within the jet, either via synchrotron radiation \citep{PSyncref}, or via secondary emission from electron-positron pairs generated in inelastic collisions between a high-energy proton and ambient matter ($pp$ interactions; \citealt{PPref}) or (internal/external) low-energy photon fields ($p$-$\gamma$ interactions, \citealt{PGammaref,PGammaref2,bottcher2007}). The $pp$ interaction channel is generally neglected for blazars because the particle density inside the emitting region is considered to be too low. According to the location of the first SED peak, blazars are further classified \citep{Blazarclasspeak} into high-frequency-peaked BL Lac objects (HBLs, peaking at X-ray frequencies) and intermediate/low-frequency-peaked BL Lac objects (IBLs/LBLs, peaking at optical-infrared frequencies). Usually the TeV-loud blazars have the first peak of the SED at UV--X-ray energies and belong to the class of HBLs. \paragraph{}

1ES 1959+650 is a well-known HBL object located nearby with a redshift $z = 0.048$ \citep{1ES1959z}. It was first detected in the radio band by the NRAO Green Bank Telescope \citep{Radiodetec} and in the X-ray band by the Einstein IPC Slew Survey \citep{Xraydetec}. Its first detection at TeV energies was by the Utah Seven Telescope Array experiment \citep{1ES1959tevdetec}. This source has also been detected in high-energy (HE; 100 MeV $<E<$ 100 GeV) $\gamma$ rays with the \textit{Fermi} Large Area Telescope (\textit{Fermi}-LAT; \citealt{3FGL}).
During May-July 2002 the source exhibited strong flaring activities and flux variations in the VHE band \citep{1ES2002aharonian,1ES2002holder,1ES2002daniel}. \citet{1ES2002krawz} performed a MWL campaign during this period including TeV $\gamma$-ray, X-ray, optical and radio observations and reported the detection of an ``orphan flare'' on 4th June 2002, a strong outburst in the VHE $\gamma$-ray band without a simultaneous X-ray counterpart. The authors reported a correlation between the X-ray and $\gamma$-ray fluxes in general except during the orphan flare. Correlated variability between these two energy bands can usually be explained by standard leptonic models whereas the lack of such a correlation challenges the SSC interpretation of the VHE flux. Hence, investigations of correlation between these two energy bands are particularly interesting for bright TeV HBLs such as 1ES 1959+650. The origin of the TeV orphan flare detected in 2002 was explained by \citet{Synchmirror2005} using a hadronic synchrotron mirror model where the flare is produced due to the interaction of relativistic protons inside the jet with external photons supplied by the reflected electron-synchrotron emission from nearby gas clouds. 
The source was later detected in a low VHE state by the MAGIC telescopes in 2004 and during the 2006 MWL campaign. The integral flux above 180 GeV is $(4.7 \pm 0.5 \pm 1.6)\times 10^{-11}\,\mathrm{cm^{2} \cdot s^{-1}}$, which is equal to $\sim 20$~\% of the Crab Nebula flux\footnote{The former error is statistical and the latter error is systematic.} above the same energy threshold~\citep{MAGIClow2004}, during the former observation. The integral flux above 300 GeV is $\sim10$~\% of the Crab Nebula flux during the latter observation~\citep{Tagliaferi2008}. 
\citet{Veritaslow2011} report the source detection by VERITAS with a significance of $16.4 \sigma$ in a low VHE flux state and higher X-ray variability compared to other energy bands. Another strong VHE outburst during 2012 was reported by \citet{Veritas2014} where an increased VHE flux was observed without a simultaneous activity in the X-ray and UV fluxes that could be explained by the reflected emission model similar to \citet{Synchmirror2005}.
Regarding the long-term behaviour in X-rays, the photon spectral index scatters below and above 2 when the spectra are fitted by a power-law (PL) function~\citep{1ES2002krawz,Patel}. This behaviour of the PL index suggests that the low-energy peak of the SED moves around the X-ray band because the photon index of a PL is $\sim2$ at the peak energy of the SED. Significant flux variabilities are also seen in HE $\gamma$ rays with the \textit{Fermi}-LAT~\citep{2015ATel.8193....1C,Patel}, although the determination of the spectral shape requires an observation longer than one day. The $\gamma$-ray spectrum in 100 MeV--100 GeV is expressed by a PL function according to the LAT 4-year point source catalogue (3FGL;~\citealt{3FGL}), and the photon power-law index is $1.88\pm 0.02$. In the 8-year point source catalogue (4FGL\footnote{https://fermi.gsfc.nasa.gov/ssc/data/access/lat/8yr_catalog/}), the power-law index is reported to be $1.82\pm 0.01$ in 50 MeV--1 TeV although a log-parabola (LogP) is preferred to a PL as the spectral type.  
The index in 10 GeV--2 TeV is $1.94\pm 0.06$ for 7-year data (the 3FHL catalogue;~\citealt{3FHL}). Therefore, the high-energy peak of the SED is anticipated to be located above 10 GeV.
\paragraph{}

The first potential association between a HE neutrino event and the flaring blazar TXS 0506+056 \citep{TXSicecube} has inaugurated a new era in multi-messenger astronomy and triggered many studies related to the neutrino-blazar coincidence \citep{TXSmagic,TXScerruti,TXSkeivani,TXSxavier}. 1ES 1959+650 is also an interesting candidate for neutrino point-source search by IceCube.  
In 2005, the AMANDA neutrino telescope reported the detection of neutrinos with a hint of spatial correlation with the source direction~\citep{Amanda2005} although the observations were not statistically significant. The most recent IceCube analysis, spanning 8 years of data however, results in a local p-value at the position of 1ES 1959+650 of $p\sim$0.25 \citep{IC_sens} which is consistent with the background-only hypothesis.\paragraph{}

Since 2015 the source entered into an active state across several energy bands, most notably in optical \citep{Opticalflare}, X-rays \citep{Xrayflare1,Xrayflare2} and also $\gamma$ rays as reported by the preliminary data analysis from the MAGIC, \textit{Fermi}-LAT, FACT and VERITAS collaborations (e.g. \citealt{2016atel1,2016atel2,2016atel3}). In this paper we report the results of a MWL campaign led by the MAGIC collaboration during April--November 2016 when the source was in an active state. The MAGIC telescopes observed three major VHE $\gamma$-ray flares from this source on 13th, 14th June and 1st July 2016. The flaring events of 1ES 1950+650 are particularly interesting because due to the close proximity and brightness of the source they allow us to perform detailed spectral measurements up to TeV energies, study the flux variability patterns, test emission models related to the origin of the VHE $\gamma$ rays and investigate their connection to cosmic-ray and neutrino production. 
Apart from 1ES 1959+650, such a detailed analysis with short-timescale variations is only possible for very bright and nearby sources such as Mrk 421 and Mrk 501 (e.g. \citealt{Mrk501_Tavecchio01,Mrk501_Ahnen18}).
The main focus of this paper is devoted to the extreme flaring events of 1ES 1959+650 from 2016, their MWL spectral and temporal properties and the investigation of their broadband characteristics to understand the underlying physical conditions inside the source during the flares. 
\paragraph{}

The paper is organised into the following sections. Section 2 describes the details of data analysis methods across all wavebands. Section 3 reports the results from the spectral and temporal analysis in the VHE band and from instruments observing at lower frequencies along with an investigation of the intra-night variability behaviour. Section 4 discusses the dimension of the emission region and broadband SED modelling of the flaring events and finally, Section 5 ends with the summary and conclusions.


\section{Observations and data analysis}
\label{sec:ObsAna}
For the present study, we performed two kinds of multi-wavelength data analysis. One is a long-term analysis of the source flux in four wavebands and the other one is a quasi-simultaneous spectral analysis at the two flux peaks, 13th and 14th June 2016. A \textit{Swift} observation was performed during the MAGIC observation on each of these days and hence we have quasi-simultaneous data from UV to VHE only for them amongst the three high VHE-flux days. 
In the following, we report a brief explanation of the instrumentation involved in these campaigns, we describe the observations performed and the data analysis techniques adopted.

\subsection{MAGIC Telescopes}
\label{ssec:ObsAna:MAGIC}
MAGIC is an array of two Imaging Atmospheric Cherenkov Telescopes (IACTs) is located at La Palma, in the Canary Islands, Spain~\citep{ALEKSIC201676}. The location coordinates are 28\degree.7N, 17\degree.9W and the altitude is 2\,200 m a.s.l. The diameter of each telescope dish is 17\,m.  The standard trigger energy threshold for low zenith angle observations is $\sim 50$ GeV~\citep{ALEKSIC201676}. 
\paragraph{}
Our data set was collected by applying two kinds of observation strategy. One is dedicated monitoring of 1ES 1959+650 in 2016. Also, intensive observations were triggered by high flux states of the source in optical, X-ray, HE $\gamma$ ray and VHE $\gamma$ ray. The effective observation time reached 72 hours over 67 nights between 29th April (MJD 57507) and 29th November (MJD 57721) 2016, including high and low flux states of the source. The majority of data from this source has been taken with a zenith angle ranging from 35\degree\ to 50\degree. For such a zenith angle, the energy threshold is higher than the value mentioned in the previous paragraph, for instance, $\sim 100$ GeV for a zenith angle of 40\degree~\citep{ALEKSIC201676}.
Some of the data were taken under moonlight.
For such data, 
the level of background noise detected by every pixel in the MAGIC cameras increases which affects the overall shape and parameters of the Cherenkov shower image. This mainly imposes non-triviality to discriminate the $\gamma$ rays from the background and to reconstruct their energy and arrival direction. Consequently, the analysis energy threshold increases. The main peculiarities and details of such data analysis are described in \citet{MAGIC-moon}. 
\paragraph{}
Observation of Cherenkov light is affected by the atmospheric transparency. The MAGIC collaboration has a self-made LIDAR system for monitoring the atmospheric transmission \citep{LIDAR2015}. For some of the observation nights, the LIDAR information was not available due to technical problems. The transmission condition can also be estimated with thermal radiation measured with a pyrometer~\citep{Martin17-AtmMonitor} and with the number of stars detected by a CCD camera, which is installed mainly for monitoring the telescope mis-pointing~\citep{Riegel05_MAGIC-tracking,Bretz09_MAGIC-drive}. 
\paragraph{}
We analysed the data using the MAGIC Standard Analysis Software (MARS; \citealt{Moralejo2009, zanin2013mars}). 
Data with aerosol transmission level from a height of 9 km above the ground level of MAGIC lower than 75\%, too high background light rate due to the moon (above 4.5 times brighter than dark conditions) and zenith angles larger than 45\degree\ were discarded to keep a low analysis energy threshold for the spectral analyses, $\sim 100$ GeV for dark conditions. After those quality cuts, $\sim 62$ hours of data over 61 nights from 29th April to 21st November survive for further analysis. For data selection based on the atmospheric transparency, the pyrometer data and the number of stars were used in addition to the LIDAR data to validate the data quality of the nights without LIDAR. For all the nights with the LIDAR data, atmospheric transmission correction based on the information was applied~\citep{LIDAR2015}.
\paragraph{}
For the long-term analysis, we derived the night-wise $\gamma$-ray flux with energies above 300 GeV. The energy threshold was set to 300 GeV for variability studies of the integral flux, to reduce its relative error. Next, we fit the observed spectrum at every night in the range 150 GeV--1 TeV with a LogP function (compatible with most of the data) to study the relation between the spectral hardness and the flux amplitude. The fitting function was folded by the energy dispersion and a model of the $\gamma$-ray absorption by extragalactic background light (EBL). Additional details about the analysis are given in Appendix~\ref{sec:Alpha-vs-Flux_details}. 

\paragraph{}
Three major flares were observed on 13th, 14th June and 1st July 2016. We performed a detailed analysis of the data during these highest VHE flux states. 
In order to reconstruct their intrinsic source spectra, the observed spectra were unfolded by the energy dispersion and the EBL absorption. In addition, we tested whether each of the following four functions can describe the intrinsic spectra: a simple PL, a PL with an exponential cutoff, a LogP and a LogP with an exponential cutoff. The explicit formulae of these functions are defined in Appendix~\ref{sec:DefSpecFunc}. 
In order to determine the peak energy of the second SED component that is constrained by the MAGIC observations, we defined two additional functions having the forms as given in Eqn.~\ref{eq:LogP_Ep} and \ref{eq:LogP-cutoff_Ep}. For these functions, the local spectral index at $E_{peak}$ was set to $-2$ in the expressions for $\frac{dF}{dE}$, thus enabling us to measure the peak location, where $E^2 \times \frac{dF}{dE}$ tends to become flat.
The fitting energy range is from 100 GeV to 9 TeV. For the spectral analyses on the three nights, we restricted the time window in order to avoid relatively strong moonlight, to get precise spectral measurements at low energy threshold.  
In addition to the spectral analyses, the intra-night variability was inspected. We produced light curves above 300 GeV with a fixed time-binning of 10 minutes for these nights and evaluated a characteristic time scale of the flux variabilities.

\subsection{\textit{Fermi} Large Area Telescope}
The LAT is one of two instruments onboard the \textit{Fermi} Gamma-ray Space Telescope \citep{LAT-2009,LAT-2012}. The LAT covers an energy range from a few tens of MeV to more than 300 GeV. It covered the whole sky every three hours during its standard survey mode.  
Its point spread function (PSF) is about 0\degree.8\ with 68\% containment at 1 GeV\footnote{http://www.slac.stanford.edu/exp/glast/groups/canda/\\lat_Performance.htm}. In order to suppress contamination of gamma rays from Galactic diffuse emission and nearby sources to 1ES 1959+650, we set the analysis range from 300 MeV to 300 GeV and used only three-fourth of the \texttt{P8R2\_SOURCE\_V6} data with better PSF, namely, the event type \texttt{PSF1}, \texttt{PSF2} and \texttt{PSF3}~\citep{Pass8-GRB}.
\paragraph{}
We performed the standard binned likelihood analysis of the data from 28th April (MJD 57506.0) to 24th November (MJD 57716.0) 2016, binning by 3 days. On top of that, we focused on the data for 1.5 days from 12th June 21:00 to 14th June 9:00 in 2016, which are quasi-simultaneous with the MAGIC data on 13th and 14th June 2016 and the central time roughly coincides with that of the MAGIC analysis period composed of those two days. In addition to the version 11-05-03 of the \textit{Fermi} standard \textit{ScienceTools}\footnote{https://fermi.gsfc.nasa.gov/ssc/data/analysis/software/v11r5p3.html}, we used the version 0.15.1 of the Fermipy python package \citep{Fermipy}.
The region of interest (RoI) for the likelihood fitting is taken to be a square of width $20 \degree$. The likelihood model includes the sources within a square region whose width is $40 \degree$. 
The included sources were taken from the LAT 4-year Point Source Catalogue (3FGL; \citealt{3FGL}). In addition, an FSRQ CGRaBS J1933+6540 (a.k.a. TXS 1933+655), which was detected by the LAT after the release of the 3FGL catalogue~\citep{LAT-detection-CGRaBSJ1933+6540}, the Galactic diffuse component and the isotropic background were also included in the model. The models for the Galactic diffuse and the isotropic component were given by the files of gll\_iem\_v06.fits and iso\_P8R2\_SOURCE\_V6\_PSF\{1-3\}\_v06.txt, respectively\footnote{https://fermi.gsfc.nasa.gov/ssc/data/access/lat/\\BackgroundModels.html}. We modelled CGRaBS J1933+6540 by a power-law spectrum and point-like spatial distribution at the position catalogued by~\citet{2002ApJS..141...13B}. The background components with the detection significance lower than $1 \sigma$ in each analysis period were removed from the model. The normalisation of the components with the significance $>3 \sigma$ and all spectral parameters of the components with the significance $>4 \sigma$ were kept free to vary for the fitting. The other parameters were fixed to the values in the 3FGL catalogue. The spectral index of 1ES 1959+650 was fixed to $1.88\pm 0.02$ in a case when the significance was lower than $4\sigma$. For the binned likelihood fit, the energy dispersion correction was enabled.

\subsection{\textit{Neil Gehrels Swift} Observatory}
 We used the publicly available data of two instruments onboard \textit{Neil Gehrels Swift} Observatory (hereafter \textit{Swift}), namely, the X-ray telescope (XRT) and the UV/Optical Telescope (UVOT) in the LAT analysis period from 28th April to 22nd November 2016, covering our MAGIC analysis period. 

\subsubsection{X-ray Telescope}
The XRT is sensitive in the energy range of 0.2--10 keV \citep{XRT2005}. The observations were performed in Window Timing mode and the exposure time distributes between 250\,s and 1\,981\,s. 1ES 1959+650 was observed with the XRT 80 times during the period.
The data were reduced with the version 0.13.3 of the standard software \texttt{xrtpipeline}. For the calibration files, the version 20160609 of the XRT \texttt{CALDB} were used. We extracted the counts within 47.1462 arcseconds from the source. 
We used the version 12.9.1 of \texttt{XSPEC} for high-level analysis. We re-binned the data to have at least 25 counts per bin and fitted the spectra only above 0.5 keV with a PL and a LogP function. For the fitting, the equivalent hydrogen column density is fixed at $10^{21}\, \mathrm{cm^{-2}}$. 
For each observation of the XRT, we performed a fit of the spectrum with \texttt{XSPEC}. 
As the long term analysis, for each observation we produced the energy-flux light curve in the energy range 0.5 to 5 keV. The results of the fit with the LogP were used also for tracing the relation between the spectral hardness and the flux. The details of the analysis are described in Appendix~\ref{sec:Alpha-vs-Flux_details}.
For the spectral analysis at the major flares, we found observations from 02:44 to 04:00, 13th (ID: 35025243) and from 02:16 to 03:20, 14th (ID: 35025245) June 2016. These X-ray times are subsets of the MAGIC observation time. The exposure time is 972 s and 865 s, respectively. We derived the differential energy flux from 0.6 keV to 7.5 keV.

\subsubsection{UV/Optical telescope}
The UVOT has six filters that have a narrow effective waveband ranging from 170 nm to 600 nm \citep{UVOT2005,UVOT2008}. We measured the energy flux in an aperture with a radius of 5\arcsec\, for one of the filters, fixing the Galactic extinction as $E(B-V) = 0.1478$ \citep{Schlafly-Finkbeiner2011}. As the background region, we took an annulus from 27.5\arcsec\, to 35\arcsec.
In order to produce a long-term light curve, we used the data of the filter W1 centred at $260$ nm. It is because the source was observed most frequently with W1 among the filters of the UVOT, 81 times from 28th April to 21st November 2016.
For the simultaneous multi-waveband analysis on 13th and 14 th June 2016, the data of the filters [W1, W2: centred at $192.8$ nm] and [W1, M1: centred at $224.6$ nm, W2] were available, respectively. 

\section{Results}
\subsection{Long-term light curves in the very-high-energy $\gamma$ ray and other wavebands} 
\begin{figure*}
  \centering
  \includegraphics[width=16cm]{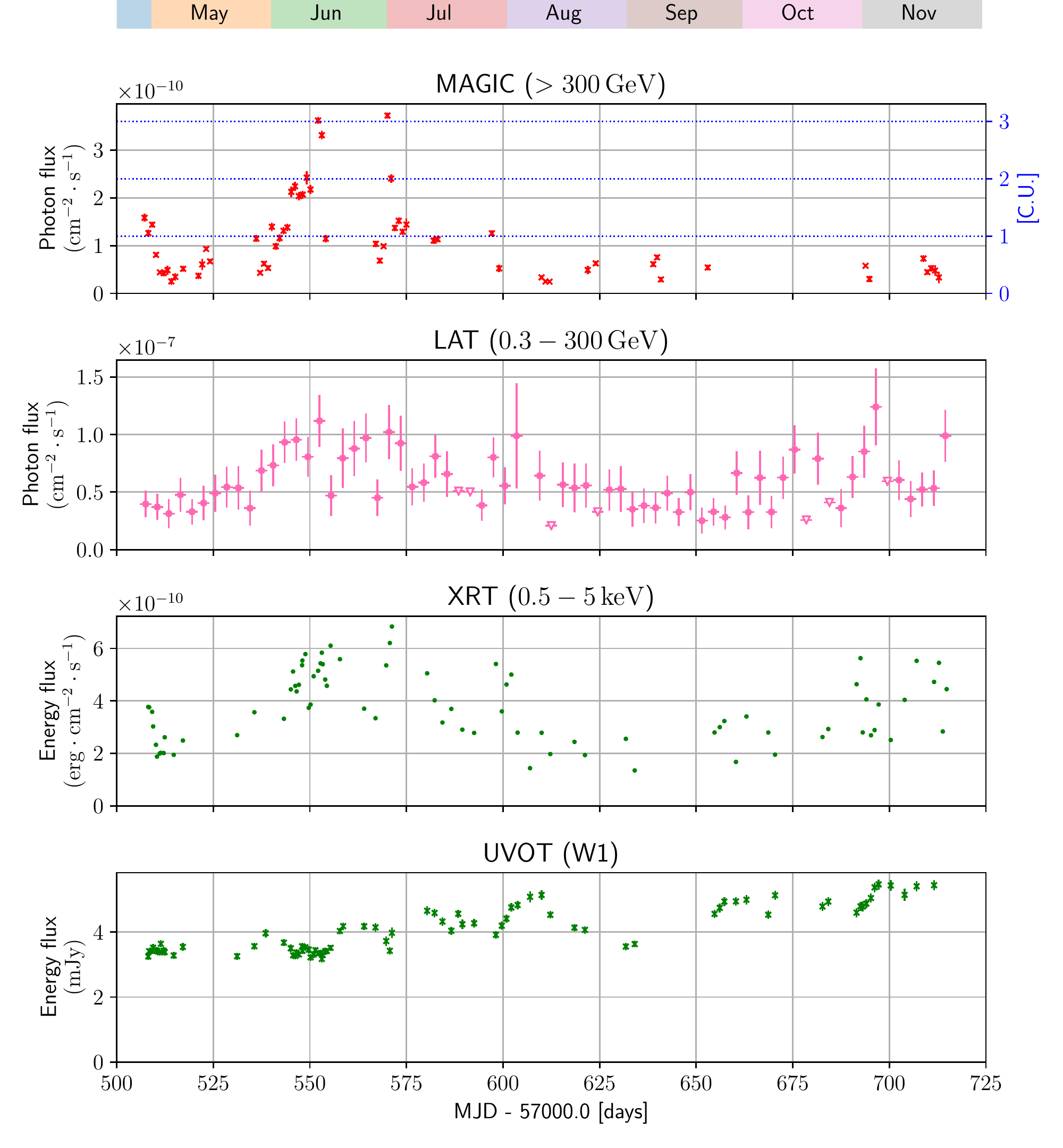}
  \caption{Long-term light curves of 1ES 1959+650 in 2016 with four instruments. From top to bottom, VHE gamma-ray flux ($>300$ GeV) from MAGIC, HE gamma-ray flux (0.3--300 GeV) measured with \textit{Fermi}-LAT, X-ray energy flux (0.5--5.0 keV) from \textit{Swift}-XRT and UV energy flux (W1 filter, $\sim 260$ nm) from \textit{Swift}-UVOT are plotted. The flux in Crab Units is indicated with an additional y-axis in the top panel. \newline (A coloured version of this figure is available in the online journal.)}
 \label{fig:LC_2016}%
\end{figure*}

The long-term flux light curve of $\gamma$ rays with energies above 300 GeV from 29th April to 21st November 2016 is displayed on the top panel of Fig.~\ref{fig:LC_2016}. It exhibits a large flux variability of more than one decade. Such an erratic trend in the light curves is a common feature of HBLs. On 13th, 14th June and 1st July 2016, the flux above 300 GeV reached $\sim 3$ Crab Units (C.U.)\footnote{flux level of the Crab Nebula under dark conditions measured above the same energy threshold~\citep{MAGIC-Crab2008}} (13th and 14th June are treated as separate flares in our work mainly due to difference in their finer-scale temporal variability; see Section~\ref{subsec:intra-night variability}). This is the highest flux level observed from this source since 2002. On the other hand, the lowest flux level is $\sim0.2$ C.U., which is comparable to the one {detected} in the past low states as mentioned in Section~\ref{sec:Introduction}. 
There have been several other smaller flares with varying rise and fall times (e.g. two flares with flux level $\sim 2$ C.U. around MJD 57550 and 57570).
The light curves obtained with the LAT, the XRT and the UVOT are plotted on the second, third and last panel of Fig.~\ref{fig:LC_2016}, respectively. The flux in the UV and HE bands changes less compared to the one observed in the X-ray and VHE bands. The relation among the flux variability in VHE, HE and X-ray looks complicated with varying rising and falling trends amongst different wavebands. 
The flux in the UV band exhibits an overall increase throughout the observation period. 

\subsection{X-ray to very-high-energy $\gamma$-ray long-term correlation}
\begin{figure*}
    \centering
    \includegraphics[width=0.8\textwidth]{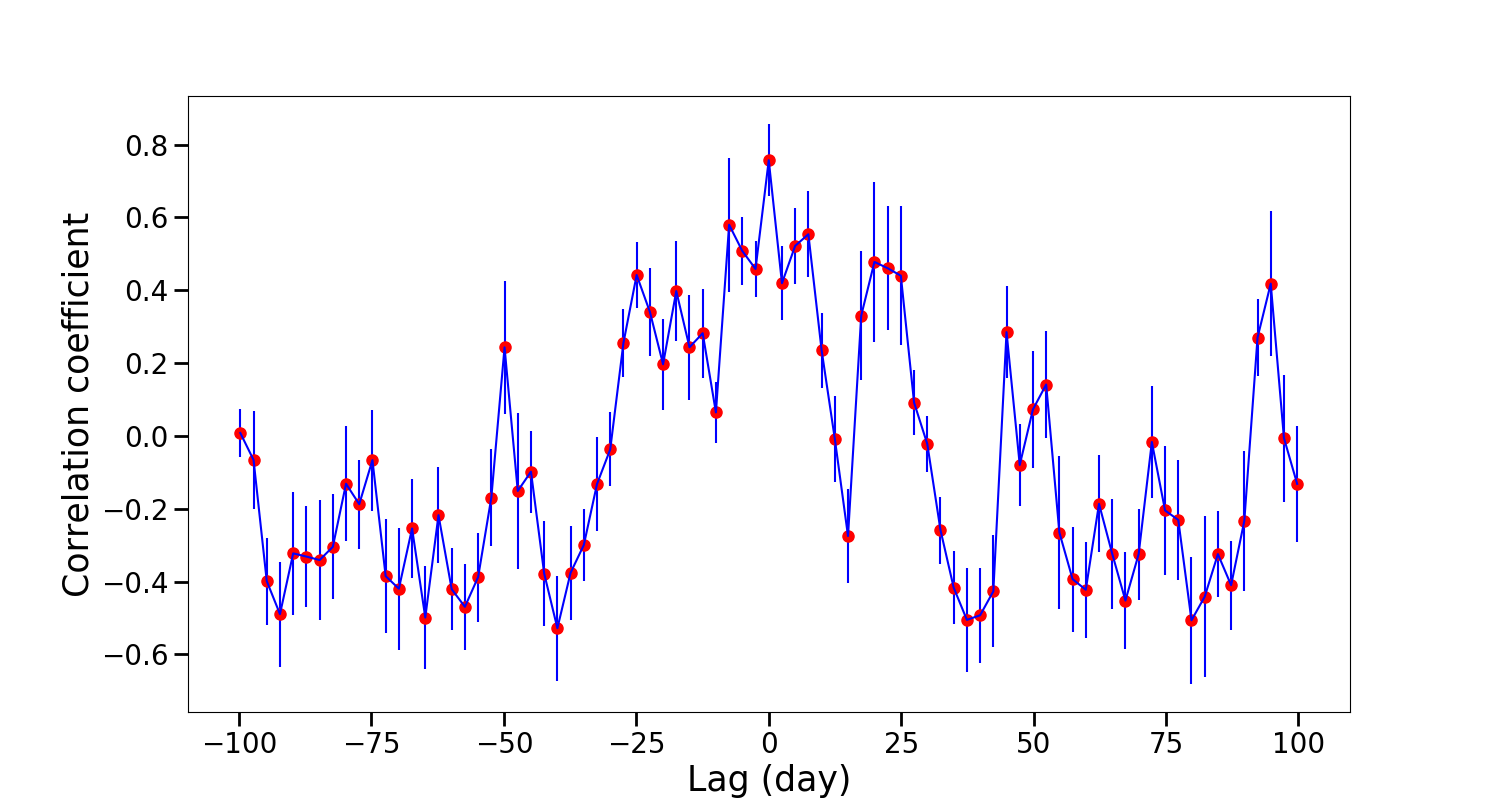}
    \caption{DCF as a function of time lag for the VHE $\gamma$-ray and X-ray light curves of the 1ES 1959+650 2016 multi-wavelength monitoring campaign. In long-term, the VHE flux shows a correlation (DCF$\sim0.76\pm0.1$) with the X-ray flux with zero time lag.}
    \label{fig:DCF_Xray_gray}
\end{figure*}

X-rays and VHE $\gamma$ rays represent the most variable energy bands of 1ES 1959+650 as shown in Fig.~\ref{fig:LC_2016}. It is anticipated that the low-energy and high-energy peaks of the SED are located in the X-ray and VHE $\gamma$-ray bands respectively and the transition between different flux states of the source mostly affect the spectra in these two bands. In order to study the correlation between these two energy bands, the method of discrete correlation function (DCF; \citealt{edelson1988discrete}) was used. Fig.~\ref{fig:DCF_Xray_gray} shows a plot of the correlation coefficients as a function of time lag in the range [-100, 100] days between X rays and VHE $\gamma$ rays. 2.5-days time-binning in lag was used to keep sufficient statistics. An overall good correlation was found in the long-term with correlation coefficient of $0.76\pm0.1$ and zero time lag, as can be seen from Fig.~\ref{fig:DCF_Xray_gray}.

It is to be noted that the long-term X-ray to VHE $\gamma$-ray correlation does not necessarily apply to the extreme flaring episodes of the source (e.g. during the 2002 MWL campaign, a general correlation was found between these two bands, except during the orphan flaring behaviour as reported in \citealt{1ES2002krawz}). For the mid-June 2016 high states of 1ES 1959+650, it is difficult to estimate the degree of correlation between X rays and VHE $\gamma$ rays due to the limited MWL statistics for the short duration flares. Hence the correlation information is unconstrained for the modelling of the flares. Our calculations can infer that the derived correlation at least holds for the long-term behaviour of the source during 2016, barring its exceptional activities.

In general, the X-ray to $\gamma$-ray cross-correlation can be quite complex with different trends between the higher and lower energy bands \citep{Mrk501_Ahnen18} or between different observation epochs. A precise cross-correlation study requires a dense long-term multi-band data set, that can provide sufficient statistics for binning the data into multiple observation periods under varying source conditions and finer energy intervals. This will be followed up in our future paper with a denser and longer multi-wavelength campaign.

\subsection{Spectral index vs. flux correlation in the very-high-energy $\gamma$ rays and X-rays}
\label{ssec:Alpha-vs-Flux}
\begin{figure*}
   \centering
    \includegraphics[width=0.8\textwidth]{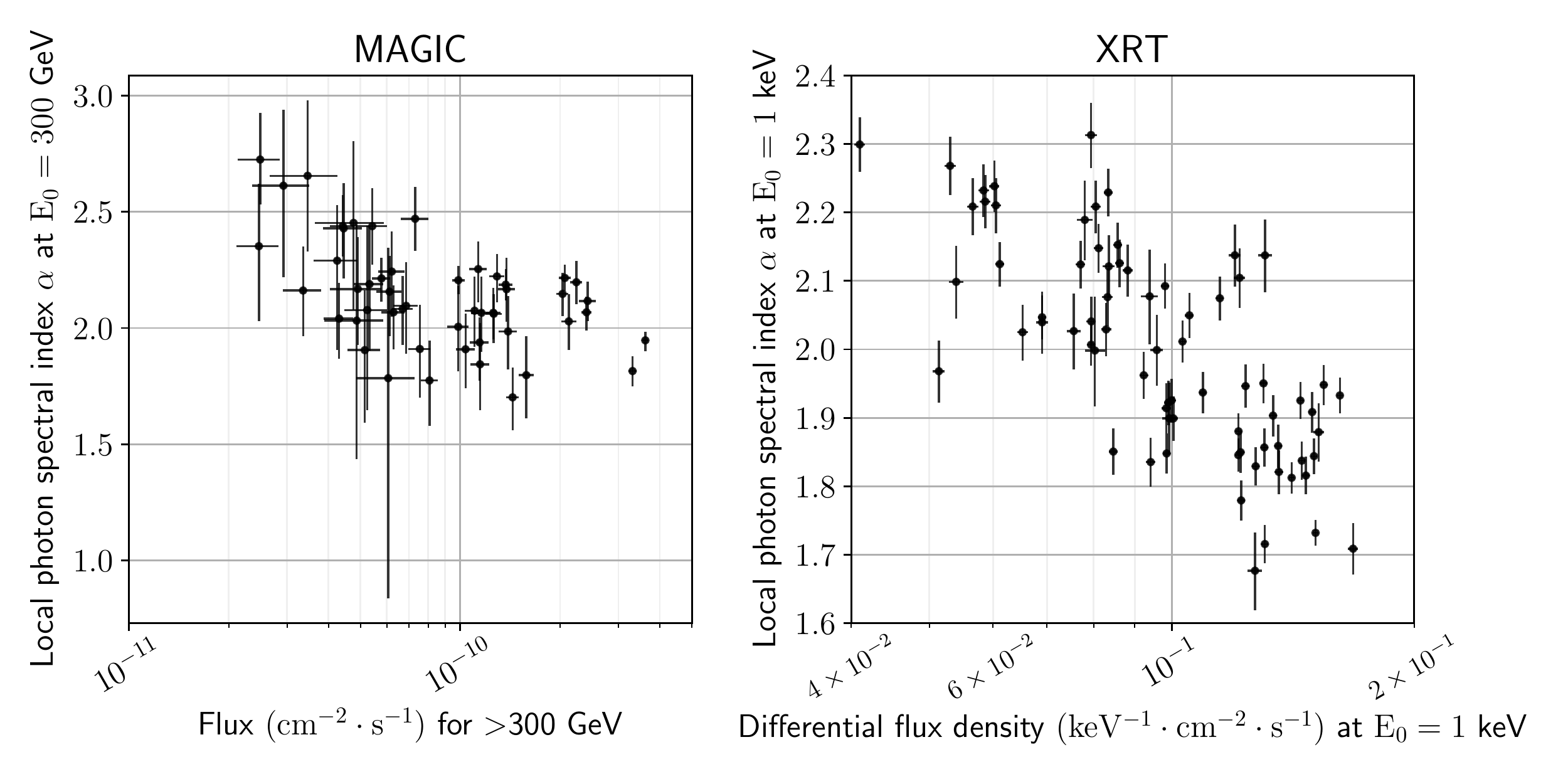}
   \caption{Correlation between $\alpha$ of LogP fitting to the MAGIC spectrum for each night (left panel) and the XRT spectrum of each observation (right). The normalisation energy is fixed at 300 GeV for MAGIC and at 1 keV for XRT. The MAGIC spectra are deabsorbed with the model of \citet{Franceschini2008}. More details can be found in the text.}
   \label{fig:indvsflux}%
\end{figure*}

In order to further investigate the long-term source behaviour in the X-ray and VHE $\gamma$-ray bands, we have studied variations of the spectral indices as a function of the source fluxes in the above two energy bands. Fig.~\ref{fig:indvsflux} shows the spectral index as a function of flux in the VHE and X-ray band respectively, for individual daily measurements with MAGIC and single observations with XRT.
The value of $\alpha$ in the LogP function given in Eqn.~\ref{eq:LogP} was used as a measure of the spectral index. It corresponds to the value of the energy-dependent PL index at the normalisation energy $\mathrm{E_0}$, which is fixed at 1 keV and 300 GeV for the XRT and the MAGIC bands respectively. As the flux value, the integrated photon flux above $\mathrm{E_0}=300$ GeV and the differential energy flux at $\mathrm{E_0}=1$ keV were used for the VHE and X-ray observations.

In order to quantify the correlation between the spectral index and the flux, we used the weighted Pearson correlation coefficient (see Appendix~\ref{Sec:Weighted Pearson correlation}). The X-ray index variation is clearly not compatible with a constant function and shows the typical harder-when-brighter behaviour throughout 2016, confirming the past trends (e.g. \citealt{1ES2002krawz}). The weighted Pearson coefficient between the index and the flux in the X-ray band is $r=0.75^{+0.05}_{-0.05}$. In the VHE $\gamma$-ray band, the correlation coefficient $r=0.64^{+0.09}_{-0.08}$, suggests a harder-when-brighter behaviour in the VHE band during 2016. A richer data set will be necessary to reinforce our claim regarding the strength of correlation. 
\paragraph{}

\subsection{Spectra during the highest flux nights }
\subsubsection{Very-high-energy $\gamma$ ray}
\begin{figure}
   \centering
   \includegraphics[width=0.48\textwidth]{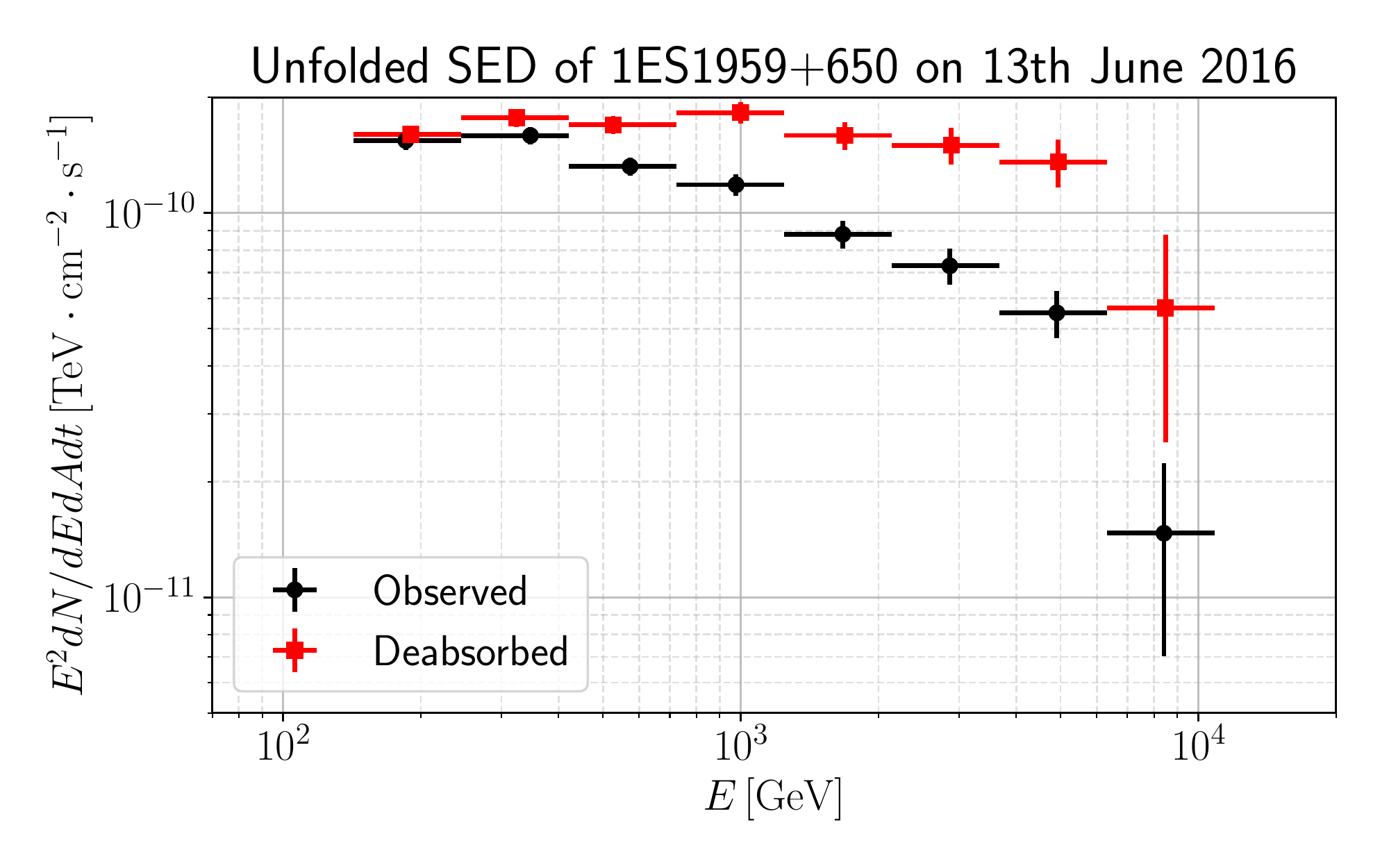}
   \includegraphics[width=0.48\textwidth]{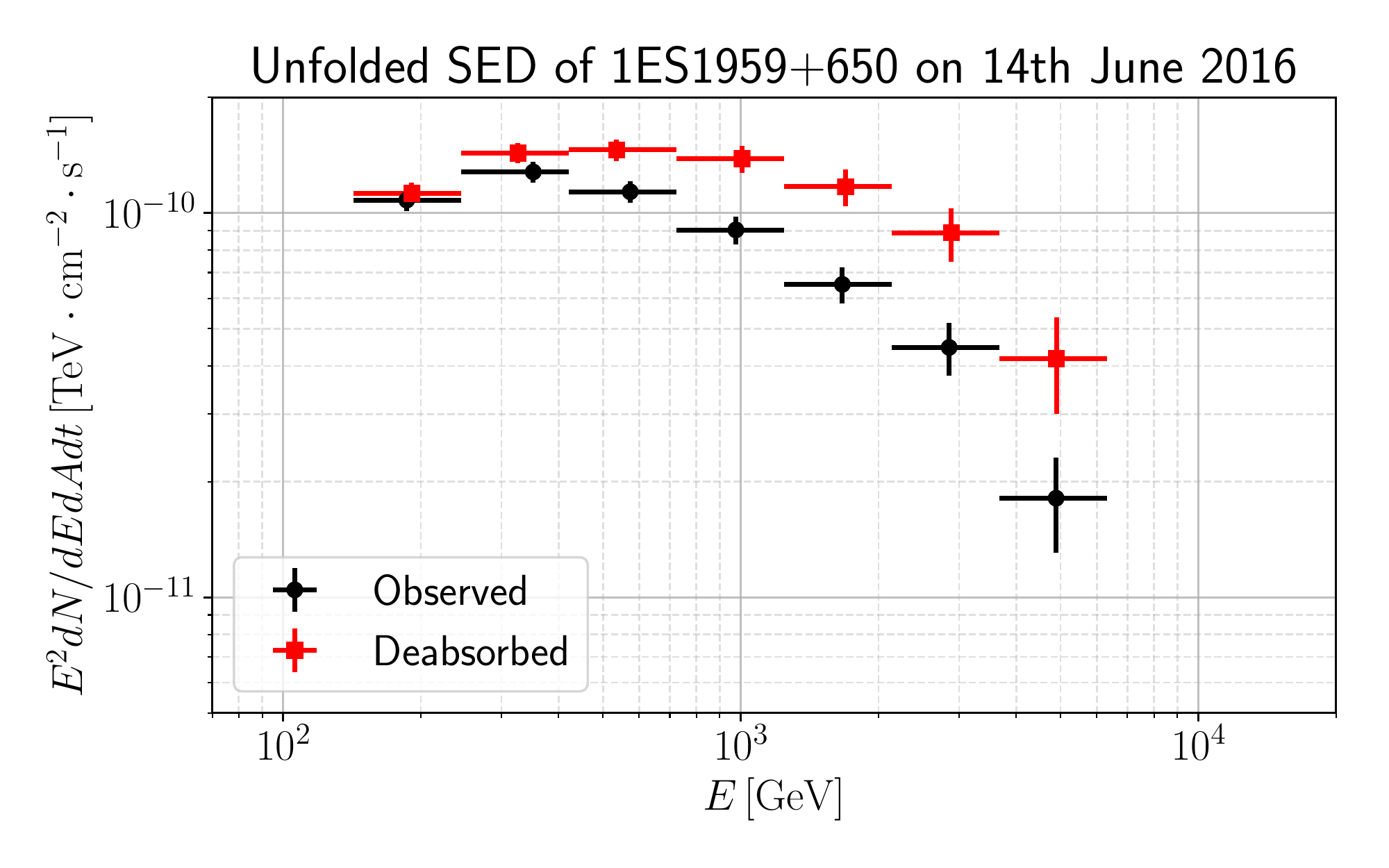}
    \includegraphics[width=0.48\textwidth]{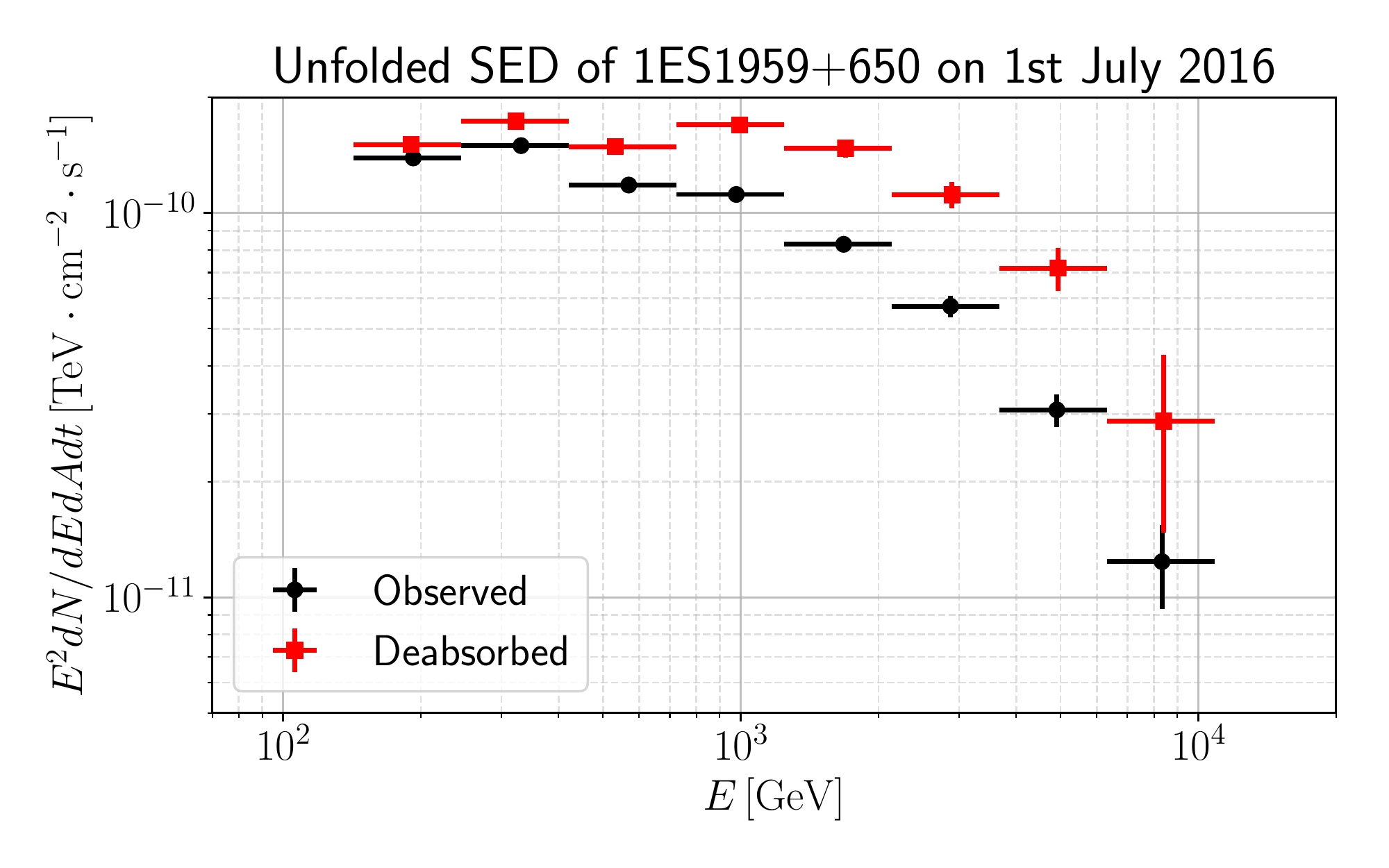}
   \caption{VHE SEDs during the highest-flux nights, 13th, 14th June and 1st July 2016 from top to bottom. The black circle and red square markers represent the observed and the EBL-deabsorbed spectra, respectively. These have been unfolded with the instrument response function of MAGIC. The absorption by the EBL has been corrected with the model of \citet{Franceschini2008}. \newline (A coloured version of this figure is available in the online journal.)}
   \label{fig:SEDs_unfolded}%
\end{figure}

A spectrum at the VHE $\gamma$ rays is especially important for this study because the high-energy peak of the SED is located in this energy range as explained in the following text and therefore it is used to constrain the spectrum of the emitting particles.
The SEDs in the VHE band on 13th, 14th June and 1st July 2016 are plotted in Fig.~\ref{fig:SEDs_unfolded}. They are unfolded with the instrument response function of MAGIC. The spectra are quite flat and extend beyond a few TeV.
We fit the SEDs during the three nights with four functions (given by Eqn.~\ref{eq:PL}--\ref{eq:LogP-cutoff}; see also Eqn.~\ref{eq:LogP_Ep} and \ref{eq:LogP-cutoff_Ep}) and the results are documented in Table~\ref{tb:SED_FitParameters}.
\begin{sidewaystable*}
\caption{Fitting parameters of the VHE spectra during the highest-flux nights in 2016.}
\label{tb:SED_FitParameters}
\centering  
\begin{tabular}{c|c|l|c|c|c|c|c|c}
Time & Flux \tablefootmark{a} & Fit model & $F_0$ & $\Gamma$ or $\alpha$ & $E_{cut}$ & $\beta$ & $E_{peak}$\tablefootmark{b} & $\chi^2/d.o.f$ \\
& $(10^{-10}\,\mathrm{cm^{-2}\cdot s^{-1})}$ & & $(10^{-9}\,\mathrm{TeV^{-1}\cdot cm^{-2}\cdot s^{-1})}$ & & (TeV) & & (TeV) & \\ \hline
13th June & {    } & \raisebox{-.45\height}{(1) PL} & \raisebox{-.45\height}{$1.81^{+0.05}_{-0.05}$} & \raisebox{-.45\height}{$2.00^{+0.02}_{-0.02}$} & \raisebox{-.45\height}{\ldots} & \raisebox{-.45\height}{\ldots} & \raisebox{-.45\height}{\ldots} & \raisebox{-.45\height}{$34.0/10$} \\[1.3ex] \cline{3-9}
02:15--04:37 & {    } & \raisebox{-.45\height}{(2) PL w/ cutoff} & \raisebox{-.45\height}{$1.93^{+0.06}_{-0.06}$} & \raisebox{-.45\height}{$1.81^{+0.05}_{-0.05}$} & \raisebox{-.45\height}{$5.4^{+1.7}_{-1.1}$} & \raisebox{-.45\height}{\ldots} & \raisebox{-.45\height}{\ldots} & \raisebox{-.45\height}{$14.1/9$} \\[1.3ex] \cline{3-9}
(MJD 57552.094& $4.06 \pm 0.13 $ & \raisebox{-.45\height}{(3) LogP} & \raisebox{-.45\height}{$1.89^{+0.05}_{-0.05}$} & \raisebox{-.45\height}{$1.83^{+0.04}_{-0.04}$} & \raisebox{-.45\height}{\ldots} & \raisebox{-.45\height}{$0.24^{+0.05}_{-0.05}$} & \raisebox{-.45\height}{$0.67^{+0.09}_{-0.07}$} & \raisebox{-.45\height}{$11.4/9$} \\[1.3ex] \cline{3-9}
--57552.192) & {    }& \raisebox{-.45\height}{(4) LogP w/ cutoff} & \raisebox{-.45\height}{$1.89^{+0.05}_{-0.05}$} & \raisebox{-.45\height}{$1.83^{+0.04}_{-0.04}$} & \raisebox{-.45\height}{$+\infty$\tablefootmark{c}} & \raisebox{-.45\height}{$0.24^{+0.05}_{-0.05}$} & \raisebox{-.45\height}{$0.67^{+0.002}_{-0.002}$} & \raisebox{-.45\height}{$11.4/8$} \\[1.3ex] \hline

14th June & {    } & \raisebox{-.45\height}{(1) PL} & \raisebox{-.45\height}{$1.46^{+0.05}_{-0.04}$} & \raisebox{-.45\height}{$2.07^{+0.03}_{-0.03}$} & \raisebox{-.45\height}{\ldots} & \raisebox{-.45\height}{\ldots} & \raisebox{-.45\height}{\ldots} & \raisebox{-.45\height}{$35.3/10$} \\[1.3ex] \cline{3-9}
02:07--03:35 & {    } & \raisebox{-.45\height}{(2) PL w/ cutoff} & \raisebox{-.45\height}{$1.67^{+0.07}_{-0.07}$} & \raisebox{-.45\height}{$1.77^{+0.07}_{-0.07}$} & \raisebox{-.45\height}{$2.9^{+0.8}_{-0.5}$} & \raisebox{-.45\height}{\ldots} & \raisebox{-.45\height}{\ldots} & \raisebox{-.45\height}{$5.9/9$} \\[1.3ex] \cline{3-9}
(MJD 57553.088& $3.28 \pm 0.13$ & \raisebox{-.45\height}{(3) LogP} & \raisebox{-.45\height}{$1.58^{+0.05}_{-0.05}$} & \raisebox{-.45\height}{$1.86^{+0.05}_{-0.05}$} & \raisebox{-.45\height}{\ldots} & \raisebox{-.45\height}{$0.36^{+0.07}_{-0.07}$} & \raisebox{-.45\height}{$0.47^{+0.05}_{-0.05}$} & \raisebox{-.45\height}{$6.0/9$} \\[1.3ex] \cline{3-9}
--57553.149) & {    } & \raisebox{-.45\height}{(4) LogP w/ cutoff} & \raisebox{-.45\height}{$1.63^{+0.09}_{-0.08}$} & \raisebox{-.45\height}{$1.81^{+0.07}_{-0.07}$} & \raisebox{-.45\height}{$5.7^{+6.2}_{-6.2}$} & \raisebox{-.45\height}{$0.18^{+0.21}_{-0.20}$} & \raisebox{-.45\height}{$1.0^{+1.8}_{-1.8}$} & \raisebox{-.45\height}{$5.3/8$} \\[1.3ex] \hline

23:59 30th June & {    }& \raisebox{-.45\height}{(1) PL} & \raisebox{-.45\height}{$1.77^{+0.03}_{-0.03}$} & \raisebox{-.45\height}{$2.10^{+0.02}_{-0.02}$} & \raisebox{-.45\height}{\ldots} & \raisebox{-.45\height}{\ldots} & \raisebox{-.45\height}{\ldots} & \raisebox{-.45\height}{$85.4/10$} \\[1.3ex] \cline{3-9}
--04:58 1st July & {    } & \raisebox{-.45\height}{(2) PL w/ cutoff} & \raisebox{-.45\height}{$1.95^{+0.04}_{-0.04}$} & \raisebox{-.45\height}{$1.86^{+0.03}_{-0.03}$} & \raisebox{-.45\height}{$3.8^{+0.6}_{-0.4}$} & \raisebox{-.45\height}{\ldots} & \raisebox{-.45\height}{\ldots} & \raisebox{-.45\height}{$11.7/9$} \\[1.3ex] \cline{3-9}
(MJD 57569.999  & $3.76 \pm 0.08$ & \raisebox{-.45\height}{(3) LogP} & \raisebox{-.45\height}{$1.87^{+0.03}_{-0.03}$} & \raisebox{-.45\height}{$1.93^{+0.03}_{-0.03}$} & \raisebox{-.45\height}{\ldots} & \raisebox{-.45\height}{$0.26^{+0.03}_{-0.03}$} & \raisebox{-.45\height}{$0.41^{+0.03}_{-0.04}$} & \raisebox{-.45\height}{$22.5/9$} \\[1.3ex] \cline{3-9}
--57570.207) & {    } & \raisebox{-.45\height}{(4) LogP w/ cutoff} & \raisebox{-.45\height}{$1.96^{+0.05}_{-0.05}$} & \raisebox{-.45\height}{$1.85^{+0.04}_{-0.04}$} & \raisebox{-.45\height}{$3.3^{+1.5}_{-0.8}$} & \raisebox{-.45\height}{$-0.05^{+0.10}_{-0.10}$} & \raisebox{-.45\height}{$+\infty$\tablefootmark{c}} & \raisebox{-.45\height}{$11.5/8$} \\[1.3ex] 
\end{tabular}
\tablefoot{The functions of PL, PL w/ cutoff, LogP, and LogP w/ cutoff are defined in Appendix~\ref{sec:DefSpecFunc} as Eqn.~\ref{eq:PL}, \ref{eq:PL-cutoff}, \ref{eq:LogP} and \ref{eq:LogP-cutoff}, respectively. The normalization energy $E_0$ is 0.3 TeV. The EBL absorption has been corrected with the model of~\cite{Franceschini2008}.}
\tablefoottext{a}{For an energy range $E> 300$ GeV.}
\tablefoottext{b}{Separate from the other parameters, only $E_{peak}$ is determined by another fitting process by expressions with $E_{peak}$, namely, Eqn.~\ref{eq:LogP_Ep}~and~\ref{eq:LogP-cutoff_Ep}.}
\tablefoottext{c}{Here $+\infty$ means that the energy is higher than the fitting range and reaches the upper limit of the parameter.}
\end{sidewaystable*}
In all cases, the EBL-corrected VHE spectra are more compatible with a curvature in the spectra rather than a simple PL (Eqn.~\ref{eq:PL}), but no decisive preference among Eqn.~\ref{eq:PL-cutoff}, \ref{eq:LogP} and \ref{eq:LogP-cutoff} were found on 13th and 14th June 2016. 
For 1st July, LogP with cutoff-type of spectrum is preferred over a pure LogP spectrum with a significance of $\sim 3 \sigma$. However, the curvature parameter $\beta$ for the LogP with cutoff spectrum is consistent with zero and the best-fit function is essentially the same as a PL with cutoff-type of spectrum.
In either case of these curved functions, the power-law index is harder than $2$ around 300 GeV.
The high flux and our intensive observations enabled us to determine the cutoff energy of the PL with a cutoff-type function and the peak energy of the LogP function to $\sim 10$\% and $\sim 20$\% statistical uncertainty, respectively. The SEDs peak at $\sim 0.4$--$0.7$ TeV and they have a cutoff above a few TeV when fitted with Eq.~\ref{eq:PL-cutoff}.
The cutoff energy and the peak energy on 13th June 2016 are higher than those on the other two nights.
These peak energies are similar to the peak of Mrk 501 SED observed in April 1997, June and July 2005~\citep{Mrk501_Djannati-Atai99,Mrk501_Albert07}. Compared to the peak energy of Mrk 501 SEDs observed at some nights in 2012~\citep{Mrk501_Ahnen18}, those peaks are a few times lower.

\subsubsection{Other wavebands} 
Now we report the results of the SED analysis for the other bands from 13th and 14th June 2016, 1st July 2016 excluded due to lack of simultaneous X-ray and optical data. The spectral points are reported in the broadband SEDs plotted in Fig.~\ref{fig:SSC_2016}, \ref{fig:psync_2016} and \ref{fig:LH_2016}.

\paragraph{}
The fitting result of the LAT spectrum for the 1.5 days is documented in Table~\ref{tb:SED_FitParameters_LAT-XRT}. The power-law index is $1.56\pm 0.20$. The index is marginally harder than the values reported in the 3FGL and the 4FGL catalogue, $1.88 \pm 0.02$ and $1.82 \pm 0.01$, respectively, by less than $2\sigma$. It should be noted that the analysis energy ranges are not identical to that of our analysis. We cannot find a significant curvature or break in the spectrum because of the small photon statistics.
The parameters of the XRT data sets, which were simultaneous with the MAGIC observations, are listed in Table~\ref{tb:SED_FitParameters_LAT-XRT}. The spectra are fitted by PL and a LogP function does not improve the goodness of the fit. The power-law index is $1.81\pm 0.01$ on 13th and $1.82 \pm 0.01$ on 14th June 2016. It is clearly harder than when the source was in a low state in 2006 as \citet{Tagliaferi2008} reported a PL index $2.197\pm 0.001$ with the data of \textit{Suzaku}-XIS for 0.7--10 keV. Our results can be compared to the SED of \citet{Tagliaferi2008} with the XIS and \textit{Suzaku}-HXD/PIN also in Fig.~\ref{fig:SSC_2016}, \ref{fig:psync_2016} and \ref{fig:LH_2016}.
\begin{table*}
\caption{Fitting parameters of the HE $\gamma$-ray and X-ray spectra from \textit{Fermi}-LAT and \textit{Swift}-XRT during the highest VHE-flux nights 2016}
\label{tb:SED_FitParameters_LAT-XRT}
\centering  
\begin{tabular}{c|c|cccc}

 & Time & Flux \tablefootmark{a} in 0.3--300 GeV & Flux\tablefootmark{b} at 1.96 GeV & $\Gamma$ & \\ 
 & & $\mathrm{(cm^{-2}\cdot s^{-1})}$ & $\mathrm{(MeV^{-1} \cdot cm^{-2}\cdot s^{-1}})$ &  &  \\ \cline{2-6} 
\textit{Fermi}-LAT & \raisebox{-.45\height}{21:00 12th--09:00 14th June} & & & & \\ 
& \raisebox{-.45\height}{(MJD 57551.875--57553.375)} & $(9.6 \pm 3.0)\times 10^{-8}$ & $(9.7 \pm 2.9)\times 10^{-12}$ & $1.56 \pm 0.20$ &  \\[1.2ex] \hline \hline

 & Time & Flux \tablefootmark{c} in 2--10 keV & Flux \tablefootmark{c} in 0.5--5 keV & $\Gamma$ &  $\chi^2/d.o.f$ \\
{   } & {    } & $\mathrm{(erg \cdot cm^{-2} \cdot s^{-1})}$ & $\mathrm{(erg \cdot cm^{-2} \cdot s^{-1})}$ & {} & {} \\ \cline{2-6} 
& \raisebox{-.45\height}{02:47--03:57 13th June} &  &   &  &  \\
\textit{Swift}-XRT & \raisebox{-.45\height}{(MJD 57552.116--57552.165)} & $4.35_{-0.07}^{+0.07}\times 10^{-10}$ & $5.15_{-0.04}^{+0.04}\times 10^{-10}$ & $1.81_{-0.01}^{+0.01}$ & $0.921$ \\[1.2ex] \cline{2-6}

& \raisebox{-.45\height}{02:18--02:33 14th June} & & & & \\
& \raisebox{-.45\height}{(MJD 57553.096--57553.106)} &  $4.88_{-0.08}^{+0.08}\times 10^{-10}$ & $5.84_{-0.05}^{+0.05}\times 10^{-10}$ & $1.82_{-0.01}^{+0.01}$ & $1.11$
\end{tabular}
\tablefoot{The fit function is PL defined in Sec.\ref{ssec:ObsAna:MAGIC}. The normalization energy is fixed at 1.96 GeV for the LAT and at 1 keV for the XRT.}
\tablefoottext{a}{Integral photon flux}
\tablefoottext{b}{Differential flux density}
\tablefoottext{c}{Integral energy flux}
\end{table*}

\subsection{Intra-night variability }\label{subsec:intra-night variability}

\begin{figure*}
   \centering
   \includegraphics[width=0.49\textwidth]{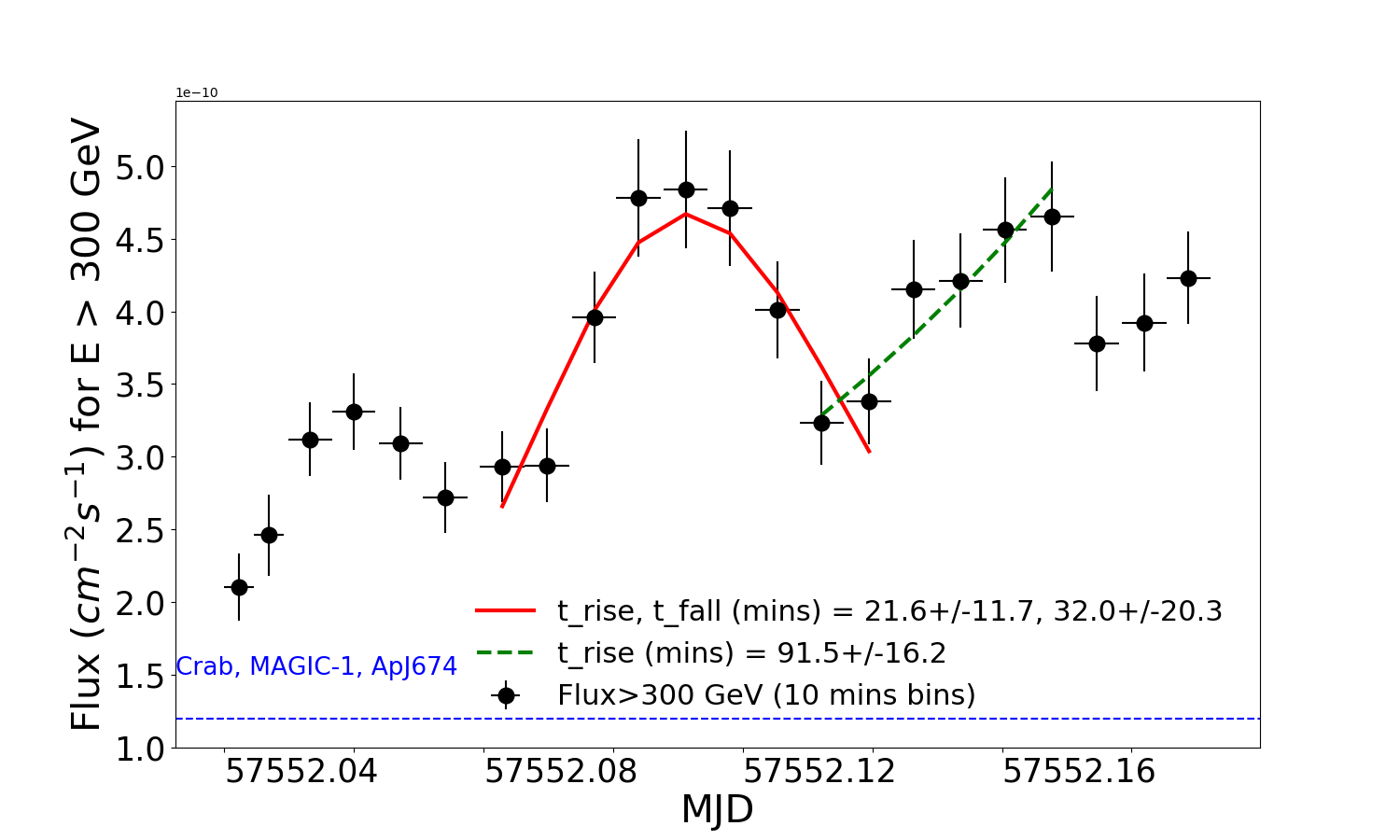}
   \includegraphics[width=0.49\textwidth]{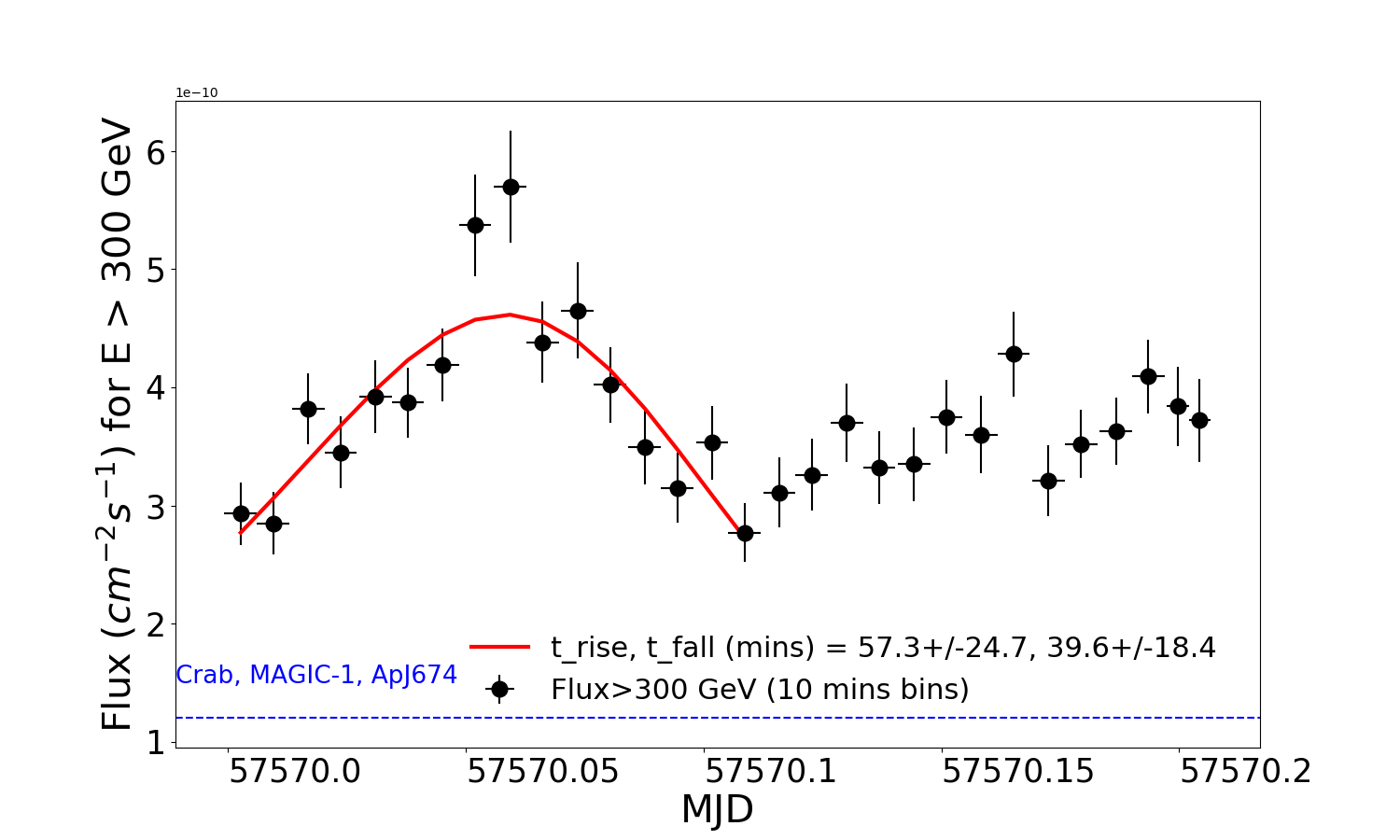}
   \caption{Fast intra-night variability observed in the VHE band on 13th June (left panel) and 1st July (right panel) 2016. The light curves above 300 GeV are constructed with a fixed time-binning of 10 minutes. Green-dashed curve: fit with the function given in Eqn.~\ref{eq:single_exp}; solid-red curve: fit with the function given in Eqn.~\ref{eq:dbl_exp}. The rise and decay times inferred from the fit are mentioned in the figure legend. The blue-dashed line represents the steady flux of the Crab Nebula above 300 GeV, shown for comparison purposes. \newline (A coloured version of this figure is available in the online journal.)}\label{fig:intra_2016}
\end{figure*}

The investigation of intra-night variability in the VHE band is not only essential to constrain the size of the emission region but also plays an important role to replicate the physical conditions inside the source leading to the origin of the second SED peak.
The observed VHE $\gamma$-ray flux exhibited fast variations for some nights in 2016, particularly for the nights with the highest VHE flux levels. We analysed the light curves with a fixed time-binning of 10 minutes and found that the light curves above 300 GeV for 13th June and 1st July 2016 show significant intra-night variability over short timescales, as shown in Fig.~\ref{fig:intra_2016}. The flux level above 300 GeV of our standard candle, the Crab Nebula is also shown for comparison purposes with a red dotted line. No significant intra-night variability was observed on 14th June 2016. A common method to quantify the mean variability of the source is given by the fractional variability amplitude \citep{Vaughan2003}. For a set of N flux points $x_i$ with corresponding errors $\sigma_{i,err}$, having mean flux $x_{mean}$ and mean squared error $\sigma_{mean, err}^2$, the fractional variability is defined with the following formula

\begin{linenomath*}
\begin{equation}
    F_{var} = \sqrt{\frac{S^2 - \sigma_{mean, err}^2}{x_{mean}^2}}
\end{equation}
\end{linenomath*}
where $S^2$ denotes the sample variance. The error in $F_{var}$ is calculated following Eqn. B2 in \citet{Vaughan2003}. The fractional variability amplitude for 13th June, 14th June and 1st July 2016 are $0.20\pm0.02, 0.06\pm0.05$ and $0.16\pm0.02$ respectively. Another approach to give a quantitative measure of the variability is to calculate the power of variability from power spectral density (PSD; \citealt{Vaughan2003}). The analysis of our data points shows that the power-law index obtained from a fit to the PSD has the hardest value for 13th June 2016, followed by
1st July. 14th June 2016 has the softest index amongst all 3 nights, which is a result similar to the one obtained from the fractional variability amplitude. However, due to a limited number of data points, determining the slope of the PSD is not very meaningful and the fractional variability amplitude gives a more reliable measure of the flux
variations. \paragraph{}

An estimate of the fastest variability timescale can be obtained from the doubling time which is defined using the following formulae from \citet{Zhang} 
\begin{linenomath*}
\begin{equation}
        t_{var, i} = \frac{F_i + F_{i+1}}{2} \frac{t_{i+1} - t_i}{|F_{i+1} - F_i|}
\end{equation}
\end{linenomath*}
where $F_i$, $F_{i+1}$ and $t_i$, $t_{i+1}$ denote the fluxes and corresponding observation times for two consecutive data points in the light curve respectively. The errors of the doubling timescale are propagated through the errors in the flux measurement. For the night of 13th June 2016, pair-wise shortest variability timescale was found between the 8th and the 9th data points having value $t_{var} = 36 \pm 14$ min. The same quantity calculated for the night of 1st July was found to have the minimum value between the 2nd and the 3rd data points with flux doubling time $t_{var} = 36 \pm 15$ min. \paragraph{}

The rise and decay time of the individual substructures in the light curves can be obtained by fitting the peaks with an exponential or sum of two exponential functions represented by the following formulae 
\begin{linenomath*}
\begin{equation}
    F(t) = A_0 e^{-|t - t_0| / t_r}
    \label{eq:single_exp}
\end{equation}
\begin{equation}
    F(t) =  A_0 / (e^{\frac{t_0 - t}{t_r}} + e^{\frac{t - t_0}{t_f}})
    \label{eq:dbl_exp}
\end{equation}
\end{linenomath*}
where $A_0$ is defined as the flux or two times the flux at $t_0$ for Eqn.~\ref{eq:single_exp} and ~\ref{eq:dbl_exp}, respectively, $t_r$, $t_f$ are the rise and decay times of the flare all of which are left as free parameters and the flux doubling time in this formalism is defined as $t_{rise/fall} = t_{r/f} \times ln(2)$. For 13th June and 1st July 2016, the results of the double-exponential fit (solid red curves in Fig.~\ref{fig:intra_2016}) and the single-exponential fit (green dashed curve in Fig.~\ref{fig:intra_2016}) are summarised in Table~\ref{tab:intranight_tab}.
The doubling timescales obtained from the fitting are comparable to the results from the \citet{Zhang} formulation, the former being slightly biased by the choice of the fitting range. For the theoretical discussions in the next section, the timescales obtained from the \citet{Zhang} formulation have been used.

\begin{table}[htb]
\caption{Results from fitting the individual substructures in the intra-night light curve of 13th June and 1st July with the functional forms given in Eqn.~\ref{eq:single_exp} and \ref{eq:dbl_exp}. $t_{rise} = t_{r} \times \log(2)$ and $t_{fall} = t_{f} \times \log(2)$}
    \centering
    \begin{tabular}{c|ccc}
    \hline
    {} & & 13th June &  \\ \hline
    
    Func. & $\chi^2 / d.o.f.$ & $t_{rise}$ \text{(min)} & $t_{fall}$ \text{(min)}  \\ \hline
    
    Single-exp fit & 1.6/3 & $91\pm16$ &\ldots \\
    
    Double-exp fit & 7.75/5 & $22\pm12$ & $32\pm20$  \\ \hline
    {} & & 1st July &  \\ \hline
    

    Double-exp fit & 19.1/12 & $57\pm25$ & $40\pm19$  \\
    \end{tabular}
    \label{tab:intranight_tab}
\end{table}

\section{Discussions}
\subsection{Size of the emission region}

The variability of blazars can act as a powerful probe to characterise the emission region and to investigate the undergoing physical processes inside the source. The emitting region is assumed to be a spherical blob of radius \textit{R} in our broadband SED models (discussed below). The observed variability timescale $t_{var}$ can be used to derive an upper limit (UL) to the size of the radiating blob in the co-moving frame of the jet from the following relation~\citep{tavecchio_variability}
\begin{linenomath*}
\begin{equation}
    R \leq \frac{c t_{var} \delta }{1 + z} \quad
\end{equation}
\end{linenomath*}
where $\delta$ represents the Doppler factor and \textit{z} represents the redshift of the source. The fastest variability timescale of the source in the VHE $\gamma$-ray band observed by MAGIC is used in our calculation for this purpose. However, we note that the spectra at the VHE $\gamma$-rays used in the broadband SED modelling described in the next section, represent an average emission state for the entire night and thus not a true representative of the finer scale variability observed in the light curve. 
Under the assumption of $t_{var} \sim$ 35 min as derived in the previous section, the upper limit to \textit{R} with $\delta = 20-60$ can be given in the range $10^{15} - 3 \times 10^{15}$ cm.  \paragraph{}

An additional constraint to the value of \textit{R} can be provided from the condition that the radius of the emission region should always be greater than the gyro-radius of the highest-energy protons, valid only for the hadronic-dominated solutions discussed below. This condition is always respected in our modelling and is given by the formula below \citep{bottcher2013}
\begin{linenomath*}
\begin{equation}
B\text{(G)} \geq 30 \frac{E_{p, \mathrm{max}}}{10^{19} \mathrm{(eV)}} \frac{10^{15}}{R \mathrm{(cm)}}
\end{equation}
\end{linenomath*}
where $E_{p, \mathrm{\mathrm{max}}}$ represents the maximum energy of the protons and \textit{B} represents the strength of the magnetic field inside the emission volume.

\subsection{Multi-wavelength spectral characteristics}
\label{ssec:MWLchar}
We have assembled quasi-simultaneous MWL data during the VHE outbursts over optical--VHE $\gamma$-ray wavebands as shown in Fig.~\ref{fig:SSC_2016}--\ref{fig:LH_2016} in order to investigate the broadband spectral behaviour of the source. For comparison, we have also plotted the MWL spectra from 2006 in Fig.~\ref{fig:SSC_2016} during a low VHE flux state measured by MAGIC (grey points; \citealt{Tagliaferi2008}). A clear shift of both the SED peaks towards higher energies (in the X-ray and VHE $\gamma$-ray domain) was observed during the flares with respect to the historical data. Also, the spectral indices are harder during the flares in both these bands. This behaviour has been reported as typical for HBLs in  \citet{Tagliaferi2008} and mentioned by many other authors in the past (e.g. \citealt{Tagliaferi2003,Pian}). The lower-energy SED peak during the flares, although not well constrained, lies above a few $10^{18}$ Hz. This indicates that the synchrotron peak frequency shifted towards the extreme-HBL (EHBL; \citealt{costamante2001_EHBL}) regime during the major flaring episodes of the source. Such a behaviour was also previously observed for Mrk 501 \citep{Mrk501_Ahnen18}. The corresponding EBL-corrected $\gamma$-ray luminosity is above $10^{45}$ erg/s which is slightly higher than the usual expectation from the luminosity-peak-frequency anti-correlation behaviour predicted in the so-called blazar sequence \citep{Blazarseq}. However, we note that the blazar sequence is constructed using average SEDs that also contain data collected during quiescent states. Moreover, the ratio between the peak luminosities of the higher and lower-energy SED components (defined by the Compton dominance parameter in the blazar sequence) changed by a factor of $\sim$4 between the historical data and the flaring $\gamma$-ray states in 2016 corresponding to  $\sim$1 order of magnitude change in the $\gamma$-ray luminosity. In the blazar sequence the Compton dominance changes by $\sim$1 order of magnitude for a $\sim$4 order of magnitude change in the luminosity.

\subsection{Broadband emission modelling}
We have modelled the broadband SEDs during the flares (13th and 14th June 2016; 1st July discarded due to lack of simultaneous X-ray and optical data) using three different theoretical frameworks considering one-zone leptonic, hadronic and lepto-hadronic models. For the modelling, a modified version of the code described in \citet{TXSmagic} was used. The emitting region is assumed to be a spherical blob of radius \textit{R} filled with a tangled magnetic field of strength \textit{B}, moving down the jet with bulk Lorentz factor $\Gamma_{bulk}$. The viewing angle of the radiated photons in the jet frame are at an angle $\theta$ with respect to an observer on Earth. The radiative output is calculated in the jet co-moving frame and then transformed to the observer frame via the Doppler factor $\delta = [\Gamma_{bulk}(1-\beta\cos(\theta))]^{-1}$. \paragraph{}

\subsubsection{Leptonic model}
\label{ssec:LeptonicModel}
\begin{figure*}
   \centering
   \includegraphics[width=0.49\textwidth]{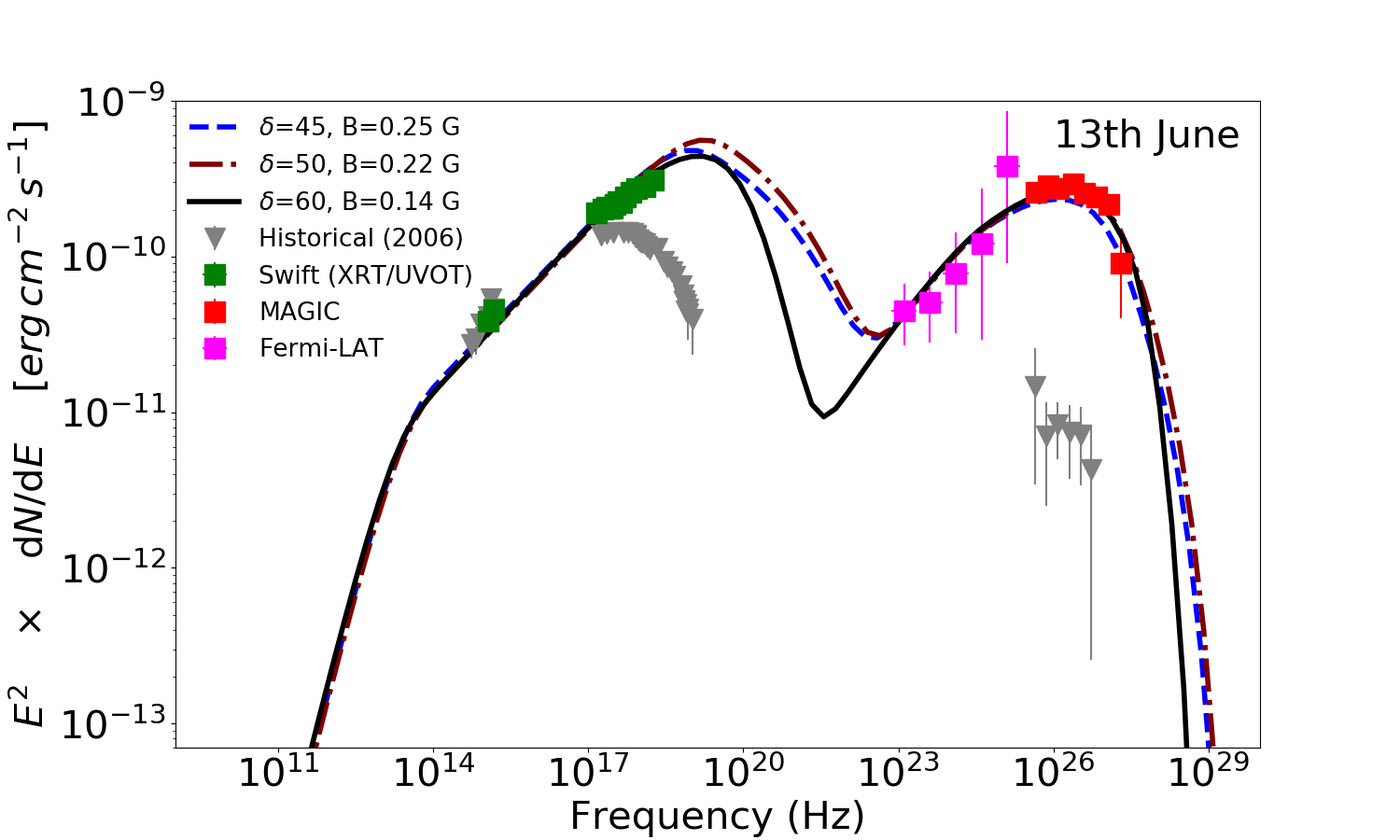}
   \includegraphics[width=0.49\textwidth]{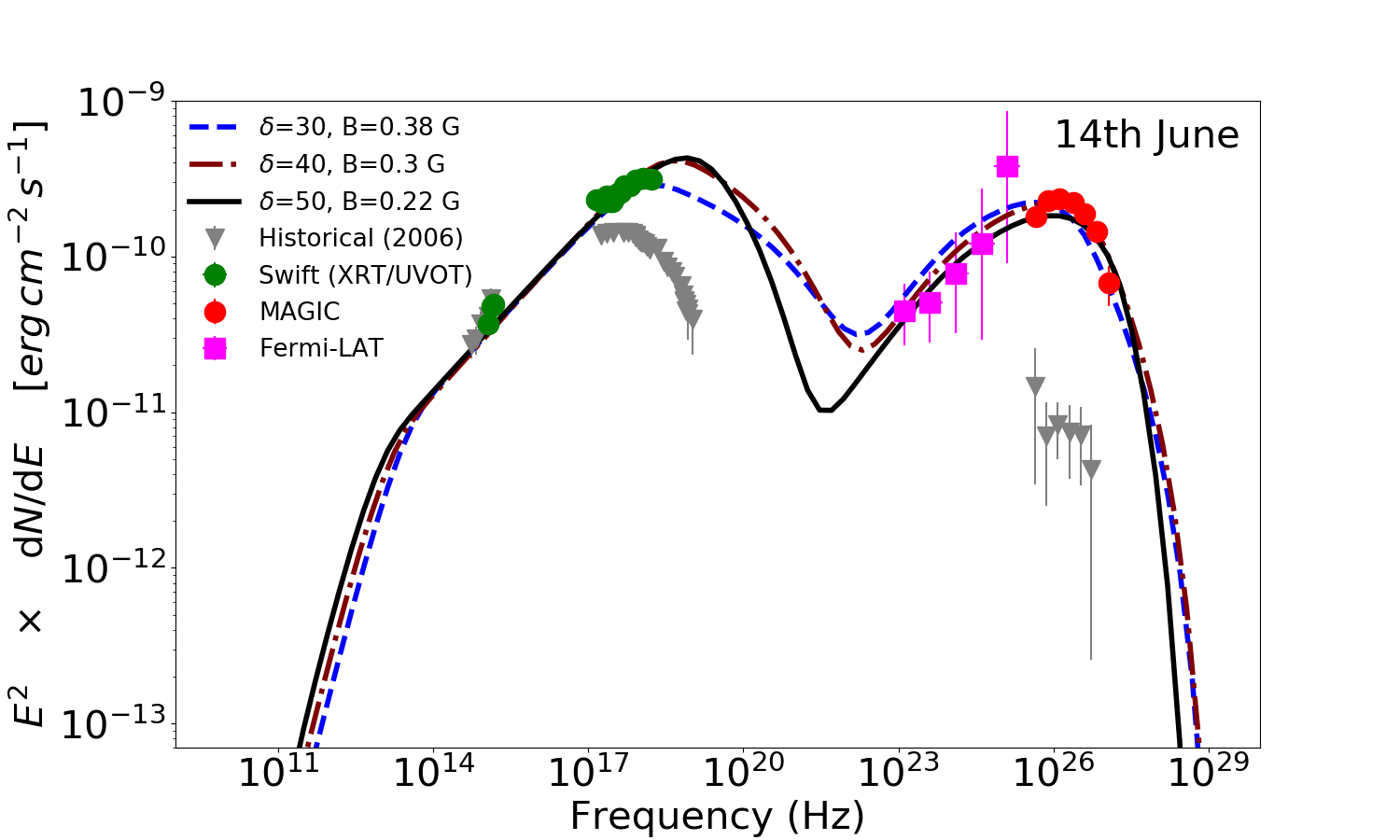}
   \caption{One-zone SSC models applied to 13th June (left panel) and 14th June (right panel) 2016. The symbols corresponding to the data sets from different instruments are given in the legend. The historical data are taken from \citet{Tagliaferi2008}. The black (solid), brown (dot-dashed) and blue (dashed) curves represent the summed emission component in increasing order of doppler factor $\delta$. We found a satisfactory explanation of the MWL data with high values of $\delta > 45$ for 13th June 2016. The data from 14th June 2016 do not strictly require such high values and can be modelled with moderate values of $\delta > 30$. For more details see the discussion in Section~\ref{ssec:LeptonicModel} and the parameters in Table~\ref{tab:SEDmodelling_params}.\newline (A coloured version of this figure is available in the online journal.)}\label{fig:SSC_2016}%
\end{figure*}

\begin{figure*}
   \centering
   \includegraphics[width=0.49\textwidth]{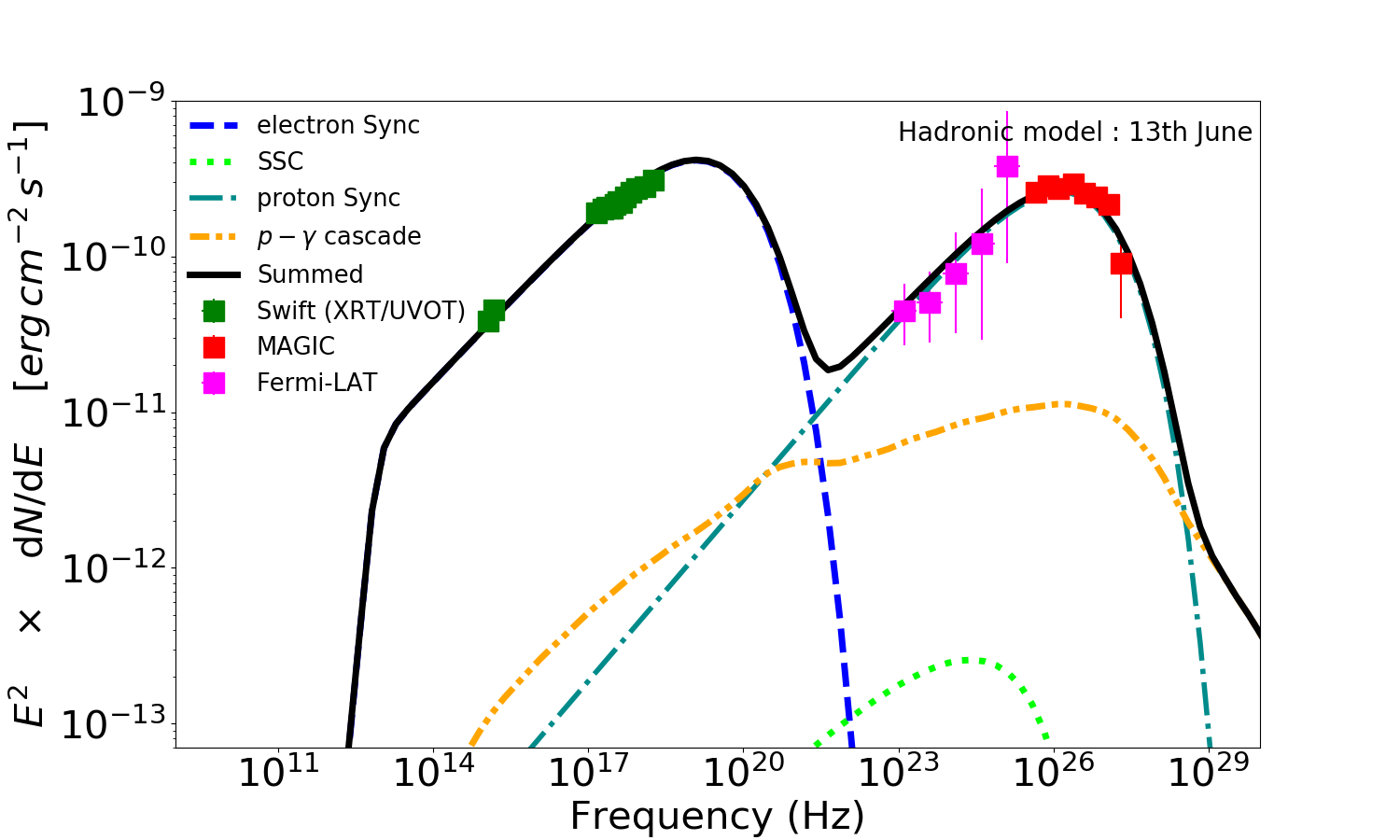}
   \includegraphics[width=0.49\textwidth]{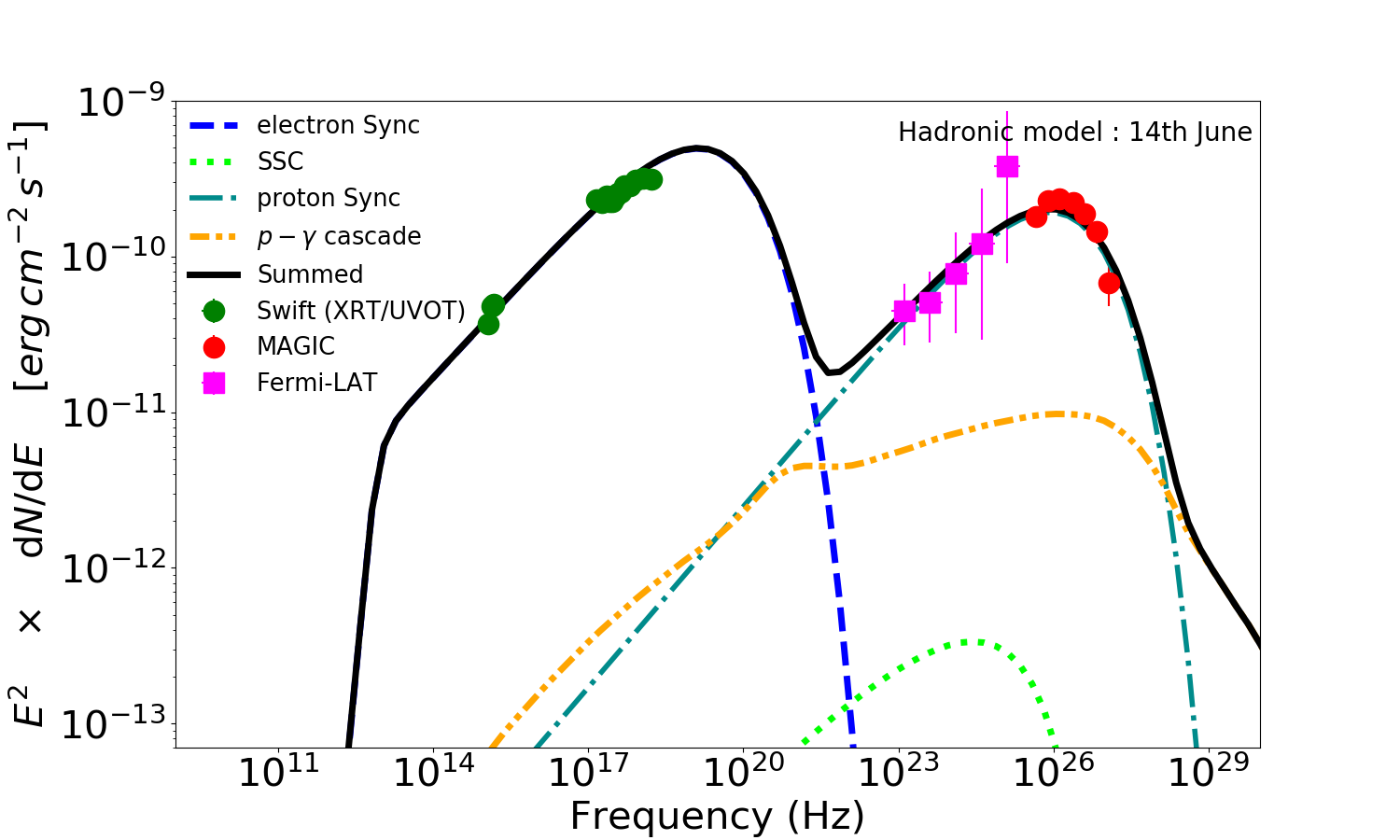}
   \caption{One-zone hadronic models applied to 13th June (left panel) and 14th June (right panel) 2016. The symbols corresponding to dataset from different instruments are given in the legend. Solid-black line: summed; dashed-blue line: electron-synchrotron; dotted-green line: SSC; dot-dashed-sea-green line: proton-synchrotron; dot-dot-dashed-orange line: $p-\gamma$ cascade. The higher energy peak in the SED is dominated by synchrotron radiation by relativistic protons which can be achieved with $B\sim 100$ G and $E_{p, \mathrm{max}} > 10^{18}$ eV and jet power $L_j \sim 10^{46}$ erg/s ($\sim L_{Edd}$). For more details see the discussion in Section~\ref{sssec:HadronicModel} and the parameters in Table~\ref{tab:SEDmodelling_params}. \newline (A coloured version of this figure is available in the online journal.)}\label{fig:psync_2016}%
\end{figure*}

First, we investigated  a one-zone SSC model for the SEDs by assuming a stationary population of primary electrons within the emitting region. The primary electron distribution is assumed to follow a broken power-law described by two slopes $n_1$, $n_2$, a break Lorentz factor $\gamma_{e, \mathrm{brk}}$, a minimum and a maximum Lorentz factor $\gamma_{e, \mathrm{min}}$, $\gamma_{e, \mathrm{\mathrm{max}}}$ respectively. The break energy is calculated by balancing the synchrotron cooling and the electron escape timescale using the following condition 
\begin{linenomath*}
\begin{equation}
    t_{e,sync} = \frac{7.75 \times 10^{8}}{B^2 \times \gamma_{e, \mathrm{brk}}} = \frac{R}{c}
    \label{eq:break_ene}
\end{equation}
\end{linenomath*}
where $t_{e,sync}$ represents the synchrotron cooling time scale of the electrons. From our modelling, values of $n_1$ between 2.25--2.3 were found to provide a satisfactory description of the \textit{Swift}-XRT and UVOT data. Under the assumption that the break in the electron spectrum is induced by radiative cooling, the second index is constrained as $n_2$ = $n_1$ + 1. Although the peak of the first SED is not well defined due to lack of simultaneous hard X-ray data, it helps to constrain the value of the magnetic field strength which is then used to derive $\gamma_{e, \mathrm{brk}}$ for a given value of \textit{R} (from Eqn.~\ref{eq:break_ene}). 
The deabsorbed VHE spectrum is quite flat and extends up to several TeV, especially for the 13th June flare.  However, due to the fast radiative cooling of the electrons and the Klein-Nishina effect, the inverse Compton component is generally suppressed and has difficulty to explain the flat photon spectrum observed at TeV energies.  To overcome this effect our model requires high values of Doppler factor and low magnetic field strength to generate the broadband spectra up to VHE in the simple SSC solutions. 
The results from the modelling are shown in Fig.~\ref{fig:SSC_2016}. The MWL SED of 13th June 2016 can be satisfactorily explained with $\delta \geq$45--50, whereas that of 14th June requires comparatively smaller values of $\delta \geq$30, which mainly arises from differences in spectral hardness/cutoff in the VHE data measured by MAGIC for 13th and 14th June 2016. Smaller values of the Doppler factor are ruled out for the range of magnetic field strength considered in this work.
The complete list of parameters for these models is reported in Table~\ref{tab:SEDmodelling_params}. 
\paragraph{}
Next, we compare these results to those of three previous flares of 1ES 1959+650 with one-zone SSC models. \citet{1ES2002krawz} applied an SSC model to the VHE high state of 1ES 1959+650 observed in 2002. The authors averaged the VHE spectra during six nights with the flux greater than 1 C.U. above 2 TeV and estimated the averaged X-ray spectrum corresponding to it. The VHE flux (here $\nu>3\times 10^{26}$ Hz) and the X-ray flux (here $\nu \sim 10^{18}$ Hz) in the averaged spectra are comparable to those in our data. The difference is roughly a few tens of percent. The authors concluded that the estimated MWL SED is reproduced by $\delta=20$, $R=5.8\times 10^{15}$ cm while other parameters are comparable to ours.
Our data set covers a much larger energy range than the one used by \citet{1ES2002krawz}, and the VHE spectra extending up to TeV energies is flatter than their model. The flat SED requires high values of the Doppler factors because such a spectrum is only reproduced by the SSC emission radiated from electrons with energy below $\gamma_{e,\mathrm{brk}}$. Since $\gamma_{e, \mathrm{brk}}$ is constrained in the previous paragraph, $\delta$ must be large so that sub-TeV $\gamma$-rays are dominantly radiated by those electrons.

In Fig.~\ref{fig:SSC_2016}, the data points taken from \citet{Tagliaferi2008} exhibit a high state in X-ray and a low state in the VHE $\gamma$-ray between 24th and 29th May 2006. The flux ratio between these two energy bands differs from that of our data roughly by a factor of $\sim$4. \citet{Tagliaferi2008} described the SED by a one-zone SSC model with $\delta=18$, $R=7.3\times 10^{15}$ cm and $B=0.25$ G. 
The difference in the luminosity ratio, which is determined by $L_{SSC}/L_{sync}=U_{sync}/U_{B}$, arises due to the difference in Compton dominance.
The increased Compton dominance in our modelling is due to a combination of 7--10 times smaller $R$ compensated by 2--3 times higher $\delta$ compared to the modelling of \citet{Tagliaferi2008}.

\citet{Veritas2014} reported a VHE flare on 20th May 2012 (MJD 56067) without simultaneously observed high X-ray state.  
The UV and X-ray spectra observed during the high and low VHE states are very similar, while at the same time being significantly different from those of the flares in June
2016.
The authors applied a time-independent SSC model to the high state and an averaged low state.
According to these models, the synchrotron peak is located at $\sim 10^{16.5}$ Hz, which is more than two orders of magnitude lower than our models. Their model peak is produced by the minimum electron Lorentz factor $\gamma_{e,min}$ of an order of $10^{6}$. This is much higher than that of our models, $\gamma_{e, min}=3$--$7\times 10^2$. 
Such high $\gamma_{e,min}$ values were also suggested by \citet{Patel} in the context of two-zone SSC models for several high-state periods and a low-state period in 2016. 
\paragraph{}
The high-energy SED peak on especially 13th June 2016 lies close to the regime of the so-called ``hard-TeV BL Lac'' objects (EHBLs with the high-energy peak above $\sim 2$ TeV).
Generally, their models require high electron spectral break energy $\gamma_{e, brk} \sim 10^6$ and large Doppler factors $\delta=20$--$60$ \citep{costamante2018_EHBL}, similar to ours. However, the magnetic field strength is extremely weak, at mG level or even lower (see e.g. \citealt{Kaufmann11_1ES0229}). 
Apart from typical EHBLs, some HBLs exhibits serendipitous EHBL-like nature temporarily. The most deeply investigated amongst them, Mrk 501 had shown harder VHE spectra or shift of the second SED peak up to $\sim$ 1 TeV during flaring activities~\citep{Mrk501_Albert07,Aliu2016_Mrk501}. The one-zone SED modelling of such states given in \citet{Mrk501_Albert07} implies [$B, \delta$]=[0.05 G, 50], [0.23 G, 25] and $R\sim 10^{15}$ cm, roughly compatible with our models.
A temporary transition of Mrk 501 towards the EHBL regime was also observed in 2012 \citep{Mrk501_Ahnen18}. 
For the strongest VHE flare on 9 June 2012, a two-zone SSC model applied by \citet{Mrk501_Ahnen18} yielded $B=6.8\times 10^{-2}$ G, $R=3.3\times 10^{15}$ cm and $\gamma_{e, min}=2\times 10^3$.
Consequently, our SSC model parameters are close to the range predicted for temporal or standard EHBLs.
This might further corroborate the transition of 1ES 1959+650 towards an EHBL-like state during the mid-June 2016 VHE outbursts.

\subsubsection{Hadronic model}
\label{sssec:HadronicModel}
We also investigated an alternative scenario, in which the high-energy component of the SED is associated with relativistic protons additionally injected into the emission region along with the primary leptons. The proton distribution is described with a power-law with an exponential cutoff function having proton spectral index $n_p$ and exponential cutoff Lorentz factor $\gamma_{p, \mathrm{\mathrm{max}}}$. We fixed the minimum proton Lorentz factor to $\gamma_{p, \mathrm{min}}$=1 in order to get a conservative estimate of the proton luminosity budget. $\gamma_{p, \mathrm{\mathrm{max}}}$ in the co-moving frame is determined from the condition $t_{acc} = minimum[t_{esc}, \,t_{psync}, \,t_{p-\gamma}]$ where $t_{acc}=10 \eta_{acc} E_p /e B c$ \citep{Cerruti2015} denotes the acceleration timescale and $t_{esc}, t_{psync}, t_{p-\gamma}$ denote the particle escape, proton-synchrotron and photo-meson cooling timescales respectively (see Fig.~\ref{fig:timescale13_2016}). The particle escape is parametrised by an efficiency factor $\eta_{esc}$ such that $t_{esc} = \eta_{esc} R / c$ \citep{Veritas2014}. 

\paragraph{}

In the hadronic scenario that we investigated, direct synchrotron radiation by the highest energy relativistic protons (few EeV in the co-moving frame) can satisfactorily reproduce the second SED peak located at few hundreds of GeV (i.e. in the VHE regime constrained by the MAGIC observations). The lower energy peak is still associated to synchrotron radiation by the primary leptons. The hadronic solutions are shown in Fig.~\ref{fig:psync_2016}. The photo-meson cascade component arises due to emission by secondary leptons that are generated when a high-energy proton interacts with the low-energy synchrotron photon field. It gives a sub-dominant contribution to the overall SED in the chosen parameter space. In the proton-synchrotron solutions, the protons have to be accelerated up to few EeV energies which can be achieved if the source possesses very high acceleration efficiency ($\eta_{acc} = 1$) under magnetically dense environments (see the timescale plots in Fig.~\ref{fig:timescale13_2016} (left panel). Large magnetic fields of the order of 100 G are adopted in our purely hadronic solutions in order to overcome the slow cooling timescale of protons which is generally insufficient to explain variability timescales of less than an hour as observed from this source during 2016. Under these conditions, the protons can cool down with timescale $t_{psync} \sim 2.5 \times 10^4$ s, shorter than the co-moving frame variability timescales ($\Delta t_{jet} = \delta \Delta t_{var}$) exhibited by the source in the VHE band. The requirement of somewhat larger values of the magnetic field is typical for proton-synchrotron models. Moreover, for our choice of R, few times $10^{14}$~cm and assuming a jet-opening angle close to 1 degree, the distance from the central core \textit{d} becomes few times $10^{16}$~cm where B $\sim$100~G can be expected (e.g. \citealt{Barkov2012}). No spectral break due to cooling is assumed in the propagated spectrum of the protons in our simple formalism since they remain un-cooled before escaping i.e. $t_{psync}$ and $t_{esc}$ are competing processes having almost equal values for the highest energy protons. In this high-\textit{B} domain, the electrons are in the fast cooling regime and parametrised by a simple power-law distribution. The complete list of model parameters for 13th and 14th June 2016 can be found in Table~\ref{tab:SEDmodelling_params}. The Doppler factor required for the hadronic solutions is considerably smaller ($\delta \sim$25) compared to that required for the purely leptonic models, especially for 13th June 2016 and represents more or less typical values. In the domain of such typical values of $\delta$, magnetic field strengths lesser than 100 G (which also implies lower values of the maximum proton energy) are insufficient to explain the flat TeV spectra of 1ES 1959+650 in purely hadronic solutions.
The difference between the VHE spectra from 13th and 14th June 2016 can be mainly attributed to small differences in the values of $\gamma_{p, \mathrm{\mathrm{max}}}$ in our hadronic solutions (14th June requires slightly smaller values of $\gamma_{p, \mathrm{\mathrm{max}}}$ than 13th June: $\sim 5 \times 10^{9}$ and $\sim 7 \times 10^{9}$ respectively). The total jet power is evaluated as
\begin{linenomath*}
\begin{equation}
    L_j = \pi R^2 c \Gamma_{bulk}^2 (u_p + u_B + u_e)
\end{equation}
\end{linenomath*}
where $u_p, u_B, u_e$ represent the energy densities carried by the protons, magnetic field and electrons respectively. In the proton-synchrotron-dominated solutions the required jet power amounts to $L_j\sim10^{46}$ erg/s, comparable to the Eddington luminosity ($L_{Edd} \sim 10^{46}$ erg/s) of the source (assuming $M_{BH} = 10^8 M_{\sun}$; \citealt{M_BH}). 
\paragraph{}
\citet{PSyncref} applied a proton-synchrotron scenario to a TeV spectrum of Mrk 501 during extraordinary flares measured in 1997. 
The EBL-corrected SED peaks at 1--2 TeV with an exponential-like cutoff around 6 TeV shared similarities to the spectrum of 1ES 1959+650 on 13th June, 2016.
The author argued that the characteristics of the TeV flares were explained by synchrotron emission from ultra-relativistic protons with $\gamma_{p.max}\geq 10^{10}$, strong magnetic field $B=30$--$100$ G and the Doppler factor $\delta=10$--$30$. These values are comparable to the parameters in our hadronic solution.\citet{PSyncref} also noted that the spectral shape is stable regardless of any possible changes in $R$ and $B$, provided $\delta$ and $\eta_{acc}$ remain unchanged. This agrees with the fact that the spectral shape in the VHE $\gamma$-ray band of the two sources is similar to each other during the flares.

\subsubsection{Lepto-hadronic model and implications for neutrino emission}
\label{sssec:LeptoHadronicModel}
 In general, the proton-synchrotron models predict neutrino fluxes below the sensitivity of the current generation of neutrino telescopes. In order to further investigate the potential of neutrino emission we also studied a lepto-hadronic model for both 13th June and 14th June 2016. The high-energy SED peak is different between the two nights mainly at the VHE $\gamma$-ray band (13th June 2016 has a slightly harder spectrum and higher VHE flux). 
 The photo-meson cascade component (the main hadronic component in our lepto-hadronic model) can take into account the differences between the VHE spectra of 13th June 2016 and 14th June 2016 with a slight difference in the particle energetics between the two nights. Our conclusions about the level of neutrino emission from the source remain the same for both nights taking into account such small differences in the $\gamma$-ray spectra. Hence we only take the night of 13th June 2016 as a reference in our paper.

\begin{figure*}
   \centering
   \includegraphics[width=0.49\textwidth]{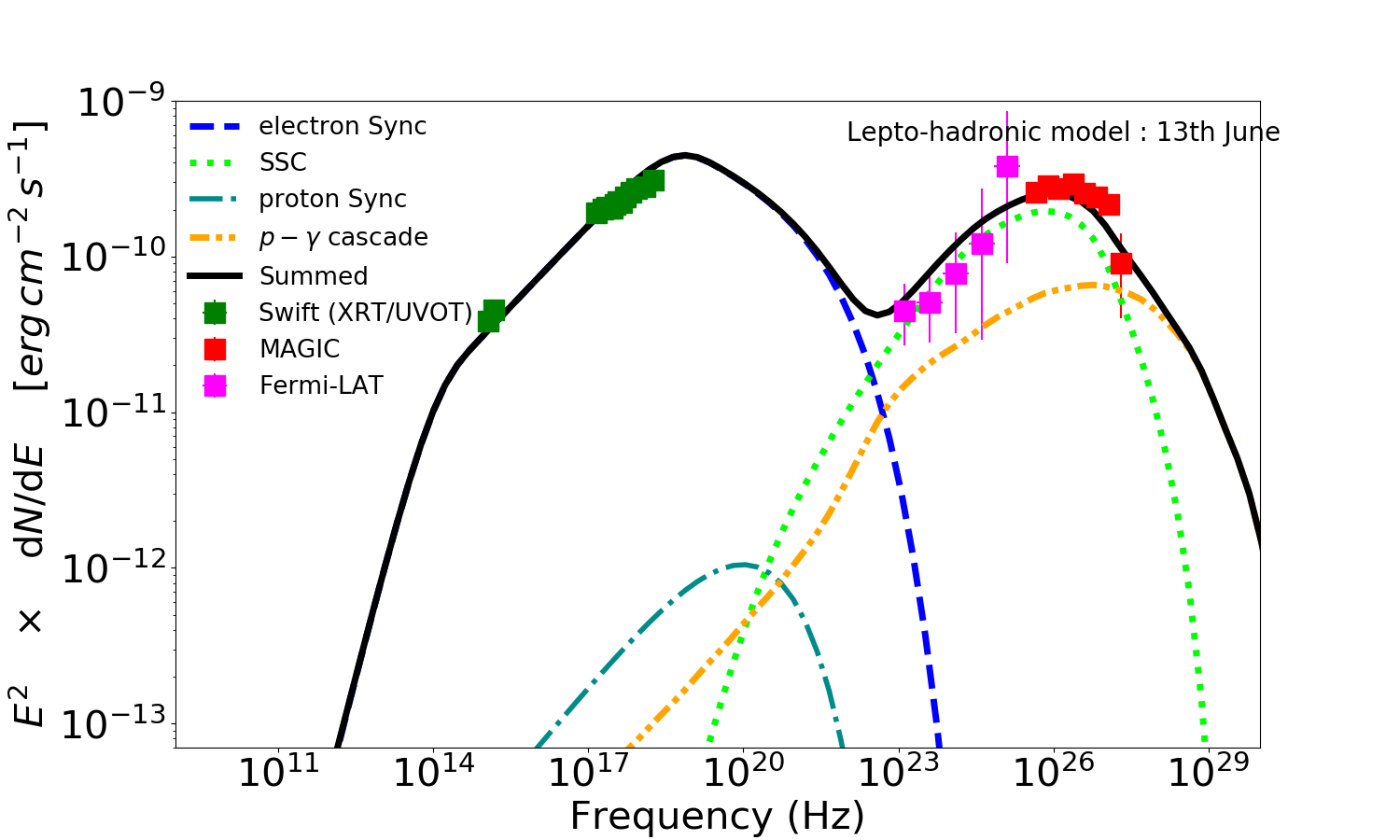}
   \includegraphics[width=0.49\textwidth]{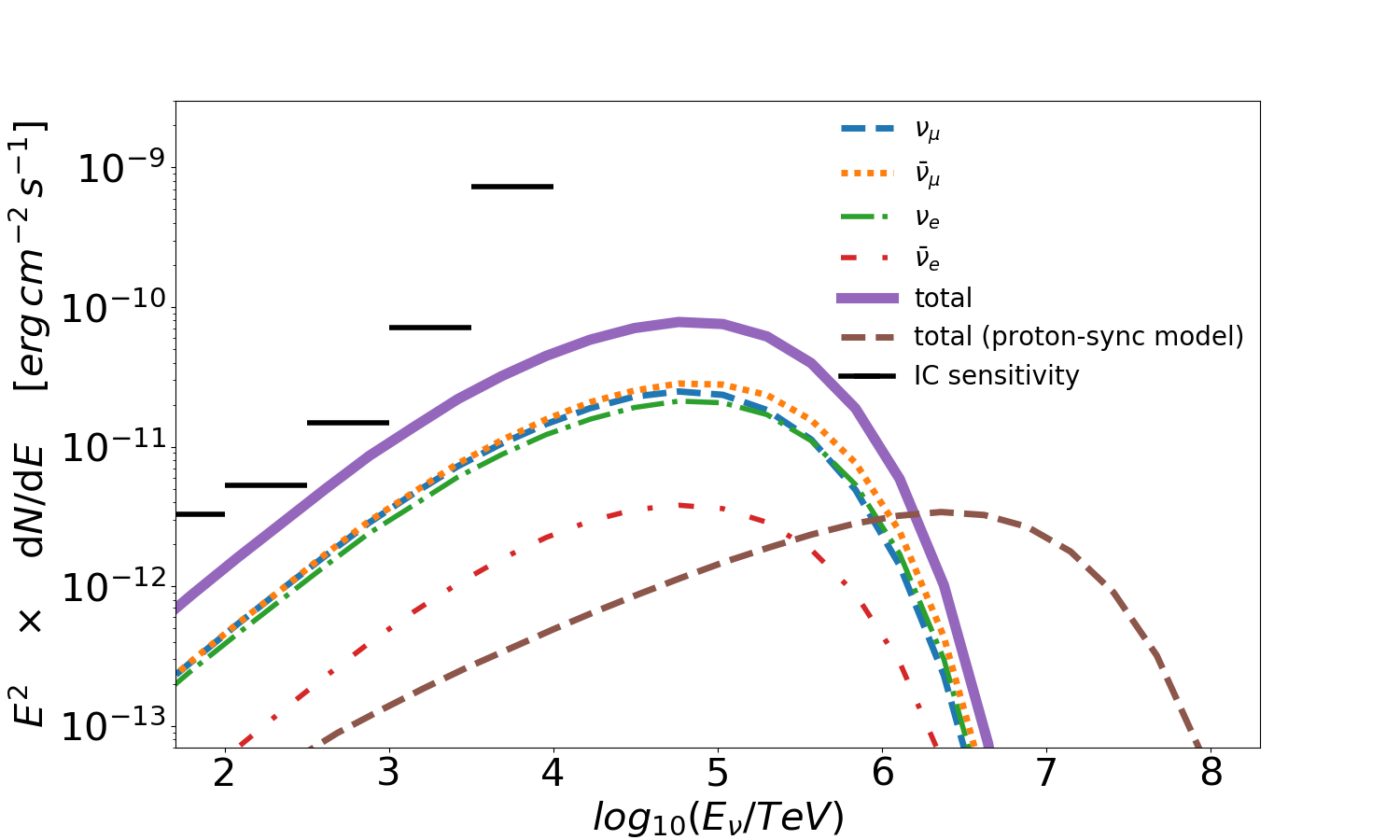}
   \caption{One-zone lepto-hadronic models (left panel) and the predicted neutrino flux (right panel) for 13th June 2016. The definition of symbols and lines in the SED model on the left panel is the same as Fig.~\ref{fig:psync_2016}. The higher energy peak in the SED in this case, is a combination of the SSC and photo-meson cascade component which can be achieved with $B\sim1$ G, $E_{p, \mathrm{max}} > 10^{16}$ eV at the cost of high jet power $L_j > 10^{48}$ erg/s ($>> L_{Edd}$). The meaning of the different curves in the neutrino spectra (right panel) is mentioned in the legend. The IceCube sensitivity curve is taken from \citet{IC_sens} corresponding to declination 60\degree. Neutrino spectra predicted in the proton-synchrotron solutions of Fig.~\ref{fig:psync_2016} peak at very high energies and provides low neutrino flux in the range 0.1--100 PeV. The neutrino peak is shifted to lower energies in the lepto-hadronic solutions providing slightly higher flux at the cost of very high values of the jet luminosity. For more details see the discussion in Section.~\ref{sssec:LeptoHadronicModel} and the parameters in Table~\ref{tab:SEDmodelling_params}. \newline (A coloured version of this figure is available in the online journal.)}\label{fig:LH_2016}%
\end{figure*}

\begin{figure*}
   \centering
   \includegraphics[width=0.49\textwidth]{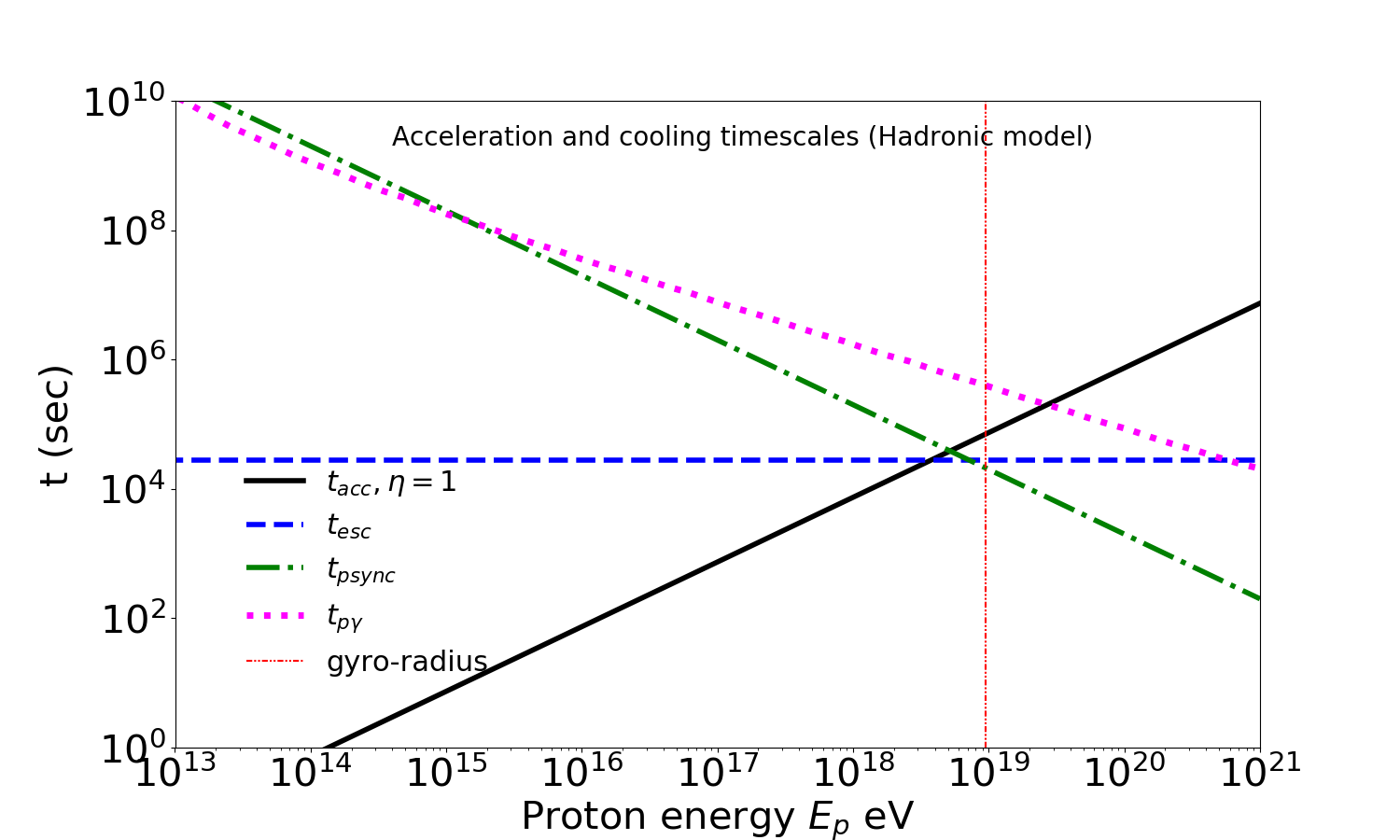}
   \includegraphics[width=0.49\textwidth]{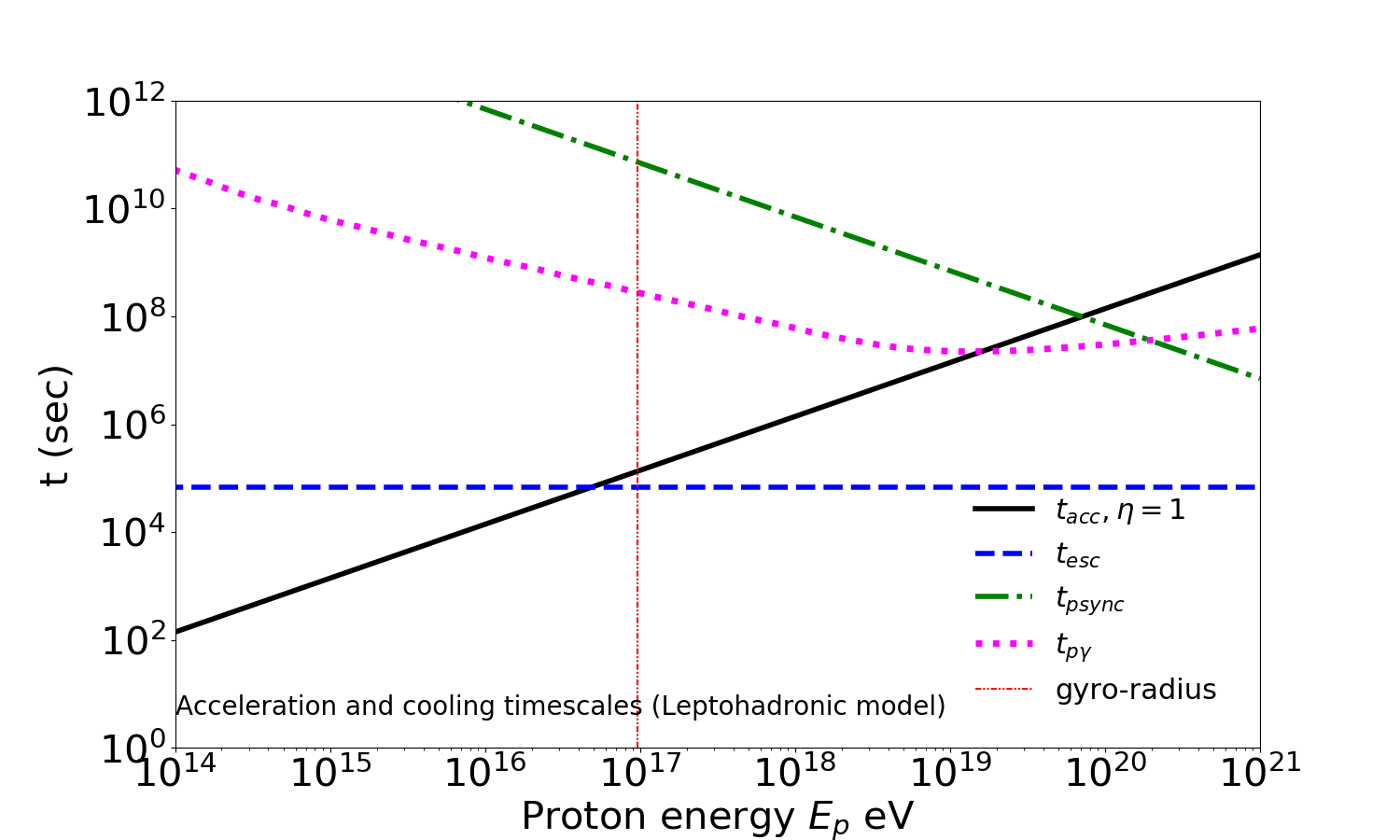}
   \caption{Comparison between acceleration timescale (solid blackline with acceleration efficiency $\eta_{acc}\,=\,1$) and different cooling timescales ($t_{esc}, t_{psync}, t_{p\gamma})$ for the hadronic (left panel) and lepto-hadronic (right panel) scenarios of 13th June 2016. Also shown is the energy at which the proton gyro-radius becomes equal to the radius of the emission region (dot-dashed-red line).\newline (A coloured version of this figure is available in the online journal.)}\label{fig:timescale13_2016}%
\end{figure*}

 We assume an additional proton population with a power-law spectrum characterised by the same functional form as described in the hadronic modelling subsection, along with the relativistic electrons inside the emission region. In the lepto-hadronic solutions, the second SED peak is comprised of contributions from both the SSC component and the $p-\gamma$ cascade component as shown in Fig.~\ref{fig:LH_2016} (left panel). In this case, the required maximum proton energy is governed by the particle escape timescale as can be seen from the timescale plot in Fig.~\ref{fig:timescale13_2016} (right panel) ($\gamma_{p,\mathrm{\mathrm{max}}} \sim 6\times10^7$). The values of the other model parameters are given in Table~\ref{tab:SEDmodelling_params}.
 
 The peak of the neutrino spectra is mainly governed by the maximum proton energy. In our proton-synchrotron solutions, due to the requirement of high values of the maximum proton energy to explain the electromagnetic SED, the inferred neutrino spectrum peaks at energies above few EeV in the observer frame and the flux at 0.1--100 PeV is quite low. In the case of the lepto-hadronic solution due to the requirement of much lower values of $\gamma_{p, \mathrm{\mathrm{max}}}$ the neutrino spectra peak about two orders of magnitude lower in energy. The inferred individual and summed components of the neutrino spectra predicted from the lepto-hadronic solution are shown in Fig.~\ref{fig:LH_2016} (right panel); for comparison, the neutrino spectrum from the proton-synchrotron model is also shown with the brown-dashed line. Additionally, the IceCube sensitivity curve, calculated for 8 years of operation and at the declination of 1ES 1959+650 \citep{IC_sens} is overlayed on the figure. Please note that a direct comparison of our model-derived neutrino spectra to the IceCube sensitivity is difficult due to variable nature of the 1ES 1959+650 electro-magnetic emission.
The neutrino spectra, calculated for a short-lasting high emission state, hardly reaches the limit of the IceCube sensitivity. Therefore one can expect that on average the neutrino emission from this source will be much lower. From the lepto-hadronic solution, the integrated neutrino flux in the range 600 GeV--100 TeV (i.e. central 90\% neutrino energy range for the declination of the source $\sim$65\degree, calculated from Fig. 1, bottom panel in \citealt{IC_sens}) is $\sim 5.5\times10^{-13} \mathrm{\,TeV\, cm^{-2}\, s^{-1}}$, which is comparable to the upper limit flux for 1ES 1959+650 obtained by IceCube ($9.86\times10^{-13} \mathrm{\,TeV \, cm^{-2} \, s^{-1}}$ at 90\% C.L.\footnote{\citealt{IC_sens}, Table 2}). Moreover, the lepto-hadronic solutions require very high values of jet power ($L_j > 10^{48}$ erg/s) exceeding $L_{Edd}$ by more than 2 orders of magnitude and hence are energetically less favourable (see however \citealt{Barkov2012}). Although relaxing the condition on the minimum proton Lorentz factor $\gamma_{p, \mathrm{min}}$=1 can reduce the luminosity to some extent, it is still insufficient to achieve sub-Eddington values. Based on the conclusions from our one-zone electromagnetic emission modelling we infer that it is difficult to produce detectable neutrino emission during the 2016 flares of 1ES 1959+650.

\begin{table*}
\caption{Parameters for the SSC, hadronic and lepto-hadronic modelling of the 13th and 14th June flares of 1ES 1959+650. 
}
    \centering
    \begin{tabular}{c|ccc|ccc}
        {} & & 13th June  & & 14th June & \\ \hline
        Parameters & SSC & Hadronic & Lepto-hadronic & SSC & Hadronic \\ \hline
        $\delta$ & 40--60 & 25 & 45 & 30--50 & 25\\
        $B$ (G) & 0.10--0.25 & 150 & 0.6 &0.2--0.4 & 150\\
        $R$ (cm) & $7\times10^{14}$--$10^{15}$ & $2.1\times10^{14}$ & $4\times10^{14}$ & $8\times10^{14}$-- $10^{15}$ & $2.1\times10^{14}$ \\
        $n_1$ & 2.2--2.3 & 2.3 & 2.3 &2.2--2.3& 2.28\\
        $n_2$ & 3.2--3.3 & {\ldots} & 3.3 &3.2--3.3& {\ldots}\\
        $\gamma_{e, \mathrm{min}}$ & $7\times10^2$ & 5 & $8\times10^2$ & $(3$--$7)\times10^2$& 5\\
        $\gamma_{e, \mathrm{max}}$ & $10^6$--$7\times10^6$ & $5\times10^4$ & $7\times10^6$ & $10^6$--$7\times10^6$ & $5\times10^4$ \\
        $\gamma_{e, \mathrm{brk}}$ & $4\times10^5$--$10^6$ & {\ldots} & $2\times10^5$ & $10^5$--$5\times10^5$ & {\ldots}\\
        $n_p$ & {\ldots} & 2.23 & 2.2 &{\ldots}&2.23\\
        $\gamma_{p, \mathrm{min}}$ & {\ldots} & 1 & 1 & {\ldots}& 1\\
        $\gamma_{p, \mathrm{max}}$ & {\ldots} & $7\times10^9$ & $6\times10^7$ & {\ldots}& $5\times10^9$ \\
        $L_j$ (erg/s) &$10^{43}$--$5\times10^{43}$  &$1.5\times10^{46}$  & $8\times10^{48}$  & $10^{43}$--$3\times10^{43}$ & $10^{46}$ \\
    \end{tabular}
    \label{tab:SEDmodelling_params}
\end{table*}

\section{Summary and conclusions}
In this work, we have reported the spectral and temporal properties of the 1ES 1959+650 MWL emission in 2016 with a special emphasis on the major VHE $\gamma$-ray flares observed by MAGIC during this period. During the 2016 long-term MWL monitoring campaign, the X-ray flux was found to be correlated in general with the VHE $\gamma$-ray flux having a discrete correlation coefficient of $0.76\pm0.1$ with no lag. For the individual extreme flaring episodes of 13th June and 14th June 2016, a correlation behaviour could not be quantified due to a lack of sufficient number of data points for the short duration flares. Hence the correlation information was not used in the broadband SED modelling. In the long-term, the X-ray spectral index hardens with increasing flux level and a hint of similar behaviour is also visible in the VHE band. In our future follow-up paper with more quasi-simultaneous multi-wavelength data, we will provide a detailed cross-correlation study between the X-ray and the VHE bands. The eventual absence of long-term correlation between these two bands might imply two different emission regions (e.g. \citealt{Patel}) or two completely independent particle populations giving rise to the emission components in the different wavebands. The blazar has shown extreme flaring episodes in the VHE band especially on 13th June, 14th June and 1st July 2016 with the highest flux $>$300 GeV reaching 2.5--3 times the Crab Nebula flux above the same energy. This is the highest flux observed from this source since the orphan flares in 2002 (the flares presented in this paper are not orphan though). The VHE spectra during the flares are quite flat extending to several TeVs. MAGIC plays a crucial role in constraining the location of the second peak in the broadband SED. The first SED peak is not well constrained due to the lack of simultaneous hard X-ray data and is treated as a somewhat free parameter in our emission modelling scenarios. \paragraph{}

The nights of 13th June and 1st July 2016 showed fast intra-night variability in the VHE band with timescales of few tens of minutes which indicates the presence of small compact emission regions inside the jet. The VHE flux on the night of 14th June 2016 was more or less constant without any signs of variations over short timescales. Owing to the different temporal characteristics of the two flares on 13th June and 14th June 2016, we consider them independent in the modelling assuming that they arise from two different independent emission zones within the jet. 
If the same emission region were to be the cause for both the flares, then in time 
$T\sim2$ days, the blob would be travelling a distance $z \sim c T \Gamma^2 \sim$1--2 parsec (assuming $\Gamma \sim$20--40). After travelling such a large distance, the blob would expand, loose energy adiabatically and its magnetic field strength would decrease \citep{Tagliaferi2008}. This would weaken the produced flux consequently, which however is not observed in the VHE band (13th June and 14th June have comparable flux levels).

\paragraph{}
We discussed the broadband spectral characteristics of the source in the framework of leptonic, hadronic and lepto-hadronic emission scenarios. In all cases we used the formulation of isotropic target photon fields for inverse Compton or photo-meson interactions, that is provided by the synchrotron radiation of the primary electrons responsible for the first SED peak. To explain the broadband spectra up to TeV energies, the SSC models require high values of Doppler factor in general ($\delta > 30$) which indicates highly-relativistic small regions in the jet responsible for the $\gamma$-ray emission. The requirement of higher $\delta$ on 13th June 2016 compared to 14th June (see Table~\ref{tab:SEDmodelling_params}) can be mainly attributed to the difference in the spectral hardness/cutoff in the VHE band. We however note, that our assumption of cooling break ($\Delta N = n_2 - n_1 =1$) comes from the simplest expectations of a break due to radiative cooling and in reality the acceleration and cooling processes can be more complex leading to a different spectral behaviour (see e.g. \citealt{Tavecchio2010}). 
In the context of this model, the unusually high flux in VHE $\gamma$-ray is considered to be produced by high Compton dominance related to the small emission region and the strong relativistic Doppler boosting compared with those of studies of different periods. In addition, some obtained parameters in the SSC model such as $B, \delta, \gamma_{e,min}$,  are similar to those predicted for Mrk 501 during an EHBL-like state. This might imply the transition of 1ES 1959+650 to such a state during extreme flaring periods.

We have also investigated alternative solutions where the jet is composed of relativistic protons in addition to the accelerated electrons. In the first scenario, the so-called proton-synchrotron model requires high values of the magnetic field strength (of the order 100~G) and acceleration efficiency close to the theoretical maximum ($\eta_{acc}\sim1$) to explain the $\gamma$-ray observations. In this parameter regime, the electrons cool down very rapidly. A hard injection spectrum of the electrons ($<2$) is thus required to explain the X-ray observations which can be generated by acceleration mechanisms such as stochastic acceleration~\citep{virtanen2005} or magnetic reconnection~\citep{cerutti2012extreme,sironi2014}.

The position of the second SED peak strongly depends on the maximum energy of the protons which in our model is determined by a balance of the acceleration and cooling timescales (proton-synchrotron, escape, photo-meson). Compared to the SSC models, the proton-synchrotron solutions require smaller values of the Doppler factor ($\delta \sim 25$). We have also investigated mixed lepto-hadronic models where the high-energy SED peak is a combination of the SSC and proton-induced cascade emission. The required jet power for the proton-synchrotron solutions is comparable to the Eddington luminosity of the source ($\sim10^{46}$ erg/s) and that for the lepto-hadronic solutions exceeds $L_{Edd}$ by about 2 orders of magnitude. However, super-Eddington values of jet power in blazars have been predicted by various other authors (e.g. \citealt{Barkov2012,Partha} etc. and the references therein). We also note that the jet power can be significantly reduced by assuming external photon fields inside the emission region as in the structured jet scenario  discussed in \citet{SL_original} (see also \citealt{Righi2017}). \paragraph{}

The neutrino spectra predicted from the proton-synchrotron model peak at very high neutrino energies (i.e. $>10^{18}$ eV in the observer frame, which is a consequence of the requirement of the high maximum proton energy in such solutions). It provides low neutrino flux in the range 0.1--100 PeV. The neutrino flux in this range can be boosted by choosing a lower value of the maximum proton energy as shown in the lepto-hadronic solutions. However, such a scenario is also energetically less favoured due to the requirement of high values of jet power as discussed above. Our predicted neutrino spectra during the brightest 2016 flare do not significantly exceed the IceCube sensitivity limit (calculated using 8 years of IceCube livetime) in all cases. The model-predicted integrated neutrino flux in the range 600 GeV--100 TeV (90\% energy confidence interval) is comparable to the flux upper limit in the location of 1ES 1959+650 derived from 8 years of IceCube data. Our conclusions are in agreement with the non-detection of significant neutrino excess in the IceCube data analysis following the 2016 $\gamma$-ray flares \citep{Kintscher2017}. 
\paragraph{}
In this work, a comparative study was done for different classes of SED models to demonstrate the multiple possibilities, which naturally leads to some degeneracy in the parameter space. Future multi-messenger and multi-wavelength observations can play a very crucial role to disentangle between the hadronic and leptonic scenarios and constrain the model parameters. For example, a multi-year multi-waveband monitoring campaign can help to follow the transition between high and low emission states. Such a long-term data set is of paramount importance to understand the nature of the emitting particles, follow the evolution of the model parameters and characterise the undergoing physical conditions which might change rapidly with the changing state of the source.  
These studies will be followed up in our future publication with a long-term monitoring campaign.

\begin{acknowledgements}
We would like to thank the Instituto de Astrof\'{\i}sica de Canarias for the excellent working conditions at the Observatorio del Roque de los Muchachos in La Palma. The financial support of the German BMBF and MPG, the Italian INFN and INAF, the Swiss National Fund SNF, the ERDF under the Spanish MINECO (FPA2015-69818-P, FPA2012-36668, FPA2015-68378-P, FPA2015-69210-C6-2-R, FPA2015-69210-C6-4-R, FPA2015-69210-C6-6-R, AYA2015-71042-P, AYA2016-76012-C3-1-P, ESP2015-71662-C2-2-P, FPA2017‐90566‐REDC), the Indian Department of Atomic Energy, the Japanese JSPS and MEXT and the Bulgarian Ministry of Education and Science, National RI Roadmap Project DO1-153/28.08.2018 is gratefully acknowledged. This work was also supported by the Spanish Centro de Excelencia ``Severo Ochoa'' SEV-2016-0588 and SEV-2015-0548, and Unidad de Excelencia ``Mar\'{\i}a de Maeztu'' MDM-2014-0369, by the Croatian Science Foundation (HrZZ) Project IP-2016-06-9782 and the University of Rijeka Project 13.12.1.3.02, by the DFG Collaborative Research Centers SFB823/C4 and SFB876/C3, the Polish National Research Centre grant UMO-2016/22/M/ST9/00382 and by the Brazilian MCTIC, CNPq and FAPERJ. 

The \textit{Fermi} LAT Collaboration acknowledges generous ongoing support
from a number of agencies and institutes that have supported both the
development and the operation of the LAT as well as scientific data analysis.
These include the National Aeronautics and Space Administration and the
Department of Energy in the United States, the Commissariat \`a l'Energie Atomique
and the Centre National de la Recherche Scientifique / Institut National de Physique
Nucl\'eaire et de Physique des Particules in France, the Agenzia Spaziale Italiana
and the Istituto Nazionale di Fisica Nucleare in Italy, the Ministry of Education,
Culture, Sports, Science and Technology (MEXT), High Energy Accelerator Research
Organization (KEK) and Japan Aerospace Exploration Agency (JAXA) in Japan, and
the K.~A.~Wallenberg Foundation, the Swedish Research Council and the
Swedish National Space Board in Sweden.
 
Additional support for science analysis during the operations phase is gratefully
acknowledged from the Istituto Nazionale di Astrofisica in Italy and the Centre
National d'\'Etudes Spatiales in France. This work performed in part under DOE
Contract DE-AC02-76SF00515.

We also appreciate a helpful framework which we used for the \textit{Fermi}-LAT data analysis, Fermipy~\citep{Fermipy}.
\end{acknowledgements}

\bibliographystyle{aa} 
\bibliography{Bibliography_MAGIC_1ES1959+650}	

\begin{appendix}
\section{Analysis details of the spectral index vs. flux correlation}
\label{sec:Alpha-vs-Flux_details}
We describe the details of the analysis reported in Section~\ref{ssec:Alpha-vs-Flux} here. 
The spectral fit is done with a LogP function given in Eqn.~\ref{eq:LogP}, which is simple and compatible with most of these spectra. For fitting the MAGIC spectra, the LogP function was folded by two functions of the photon energy. One is the dispersion of the reconstructed energy from the true value, and the other one is a correction factor for photon absorption due to the EBL (using the model of \citealt{Franceschini2008}). On the other hand, for reconstructing the intrinsic source spectra shown in Fig.~\ref{fig:SEDs_unfolded}, we unfolded the observed spectra by the energy dispersion and the EBL absorption. 
\paragraph{}
The following procedure is common for the MAGIC and XRT analyses.
The energy-dependent photon index is defined as $\Gamma(E) = \alpha + 2 \beta \log_{10} (E / E_0)$ (see Eqn. 4 in \citealt{logpindex}). The fitting range of the VHE spectra is restricted to 150 GeV--1 TeV in order to avoid a possible high-energy cutoff. For each fit, the value of $\chi^2$ is calculated. If the LogP function deviates from the spectrum so that the fit probability is smaller than 5\%, the night (for MAGIC) or observation (for XRT) is removed from the sample. The cut by 5\% corresponds to about $2 \sigma$ and leaves most of the data points compatible with the LogP spectral shape. For the nights/observations that satisfy the above criteria, we adopted the value of $\alpha$ as a measure of the local spectral index at the normalization energy $\mathrm{E_0}$.

\section{Spectral fitting functions}
\label{sec:DefSpecFunc}
The functions which were used for the spectral fitting are defined as follows:
a simple power-law (PL) 
\begin{linenomath*}
\begin{equation}
\label{eq:PL}
\frac{\mathrm{d}F}{\mathrm{d}E} = F_0 \left(\frac{E}{\mathrm{E_0}}\right)^{-\Gamma},
\end{equation}
a PL with an exponential cutoff
\begin{equation}
\label{eq:PL-cutoff}
\frac{\mathrm{d}F}{\mathrm{d}E} = F_0 \left(\frac{E}{\mathrm{E_0}}\right)^{-\Gamma} \exp\left(-\frac{E}{E_{cut}}\right),
\end{equation}
a log-parabola (LogP)
\begin{equation}
\label{eq:LogP}
\frac{\mathrm{d}F}{\mathrm{d}E} = F_0 \left(\frac{E}{\mathrm{E_0}}\right)^{-\alpha-\beta [\log_{10}(E/\mathrm{E_0})]},
\end{equation}
and a LogP with an exponential cutoff 
\begin{equation}
\label{eq:LogP-cutoff}
\frac{\mathrm{d}F}{\mathrm{d}E} = F_0 \left(\frac{E}{\mathrm{E_0}}\right)^{-\alpha-\beta [\log_{10}(E/\mathrm{E_0})]}\exp\left(-\frac{E}{E_{cut}}\right),
\end{equation}
where $\mathrm{d}F/\mathrm{d}E$ is the differential $\gamma$-ray flux as a function of the energy $E$. The value of the normalization energy $\mathrm{E_{0}}$ is fixed at 300 GeV. The expressions of the LogP and the LogP with a cutoff with $E_{peak}$ are 
\begin{equation}
\frac{\mathrm{d}F}{\mathrm{d}E} = F_0 \left(\frac{E}{E_{peak}}\right)^{-2} 10^{-\beta [\log(E/E_{peak})]^{2}}  
\tag{\ref{eq:LogP}\textprime} \label{eq:LogP_Ep}
\end{equation}
and
\begin{equation}
\frac{\mathrm{d}F}{\mathrm{d}E} = F_0 \left(\frac{E}{E_{peak}}\right)^{-2} 10^{-\beta [\log_{10}(E/E_{peak})]^{2}} \exp\left(-\frac{E}{E_{cut}}\right),
\tag{\ref{eq:LogP-cutoff}\textprime} \label{eq:LogP-cutoff_Ep}
\end{equation}
\end{linenomath*}
respectively.
The parameter $E_{peak}$ corresponds to the peak energy of a SED.

\section{Weighted Pearson correlation coefficient calculation}
\label{Sec:Weighted Pearson correlation}
We calculate the weighted mean and weighted covariance for two quantities $x$ and $y$ (in our case spectral index and flux) with errors $\sigma_x$, $\sigma_y$ respectively using the following formulae
\begin{equation}
    \mathrm{mean}(x; w) = \frac{\Sigma_i \, w_i x_i}{\Sigma_i \, w_i} \quad \mathrm{(similar \, expression \, for \, y)}
\end{equation}
where $w_i = 1 / \sigma_{x\mathbf{i}}^2$ or $w_i = 1 / \sigma_{y\mathbf{i}}^2$.
\begin{equation}
    \mathrm{cov}(x, y ; w) = \frac{\Sigma_i \, w_i \, (x_i - \mathrm{mean}(x; w)).(y_i - \mathrm{mean}(y; w))}{\Sigma_i \, w_i}
\end{equation}
where $w_i = 1 / (\sigma_{x\mathbf{i}} \sigma_{y\mathbf{i}})$.

Using the above definitions, the weighted Pearson correlation coefficient can be calculated as
\begin{equation}
    \mathrm{corr}(x,y; w) = \frac{\mathrm{cov}(x, y; w)}{\sqrt{\mathrm{cov}(x,x;w). \mathrm{cov}(y,y;w)}}
\end{equation}
The errors of the correlation coefficient have been calculated using the z-transformed Discrete Correlation Function (\citealt{Corrcoeferror}; see also \citealt{edelson1988discrete} and \citealt{Corrcoeferror_magic}).
\end{appendix}
\end{document}